\documentclass[12pt]{article}
\usepackage{epsf,fullpage}
\begin{document}

%
%
\newcommand\eqref[1]{(\ref{#1})}
\newcommand\beq{\begin{equation}}
\newcommand\eeq{\end{equation}}
\newcommand\bea{\begin{eqnarray}}
\newcommand\eea{\end{eqnarray}}
\newcommand\beano{\begin{eqnarray*}}
\newcommand\eeano{\end{eqnarray*}}
\newcommand\ben{\begin{enumerate}}
\newcommand\een{\end{enumerate}}

\newcommand\cf{{\it cf.\ }}
\newcommand\eg{{\it e.g.},\ }
\newcommand\Eg{{\it E.g.},\ }
\newcommand\ie{{\it i.e.},\ }
\newcommand\Ie{{\it I.e.},\ }
\newcommand\etal{{\it et al.\ }}
\newcommand\etalc{{\it et al.},\ }
\newcommand\etc{{\it etc.\ }}
\newcommand\etcc{{\it etc.},\ }
\newcommand\via{{\it via\ }}
\newcommand\viz{{\it viz.},\ }
\newcommand\al{\alpha}
\newcommand\be{{\beta}}
\newcommand\ga{{\gamma}}
\newcommand\de{{\delta}}
\newcommand\eps{\epsilon}
\newcommand\ka{{\kappa}}
\newcommand\la{\lambda}
\newcommand\om{{\omega}}
\newcommand\si{{\sigma}}
\renewcommand\th{{\theta}}
\newcommand\D{{\Delta}}
\newcommand\Ga{{\Gamma}}
\newcommand\La{{\Lambda}}
\newcommand\Si{{\Sigma}}
\newcommand\Om{{\Omega}}
\newcommand\mnrs{{\mu\nu\rho\sigma}}
\newcommand\mnrst{{\mu\nu\rho\sigma\tau}}
\newcommand\munu{{\mu\nu}}
\renewcommand\AA{{\cal A}}
\newcommand\CC{{\cal C}}
\newcommand\DD{{\cal D}}
\newcommand\HH{{\cal H}}
\newcommand\II{{\cal I}}
\newcommand\LL{{\cal L}}
\newcommand\MM{{\cal M}}
\newcommand\NN{{\cal N}}
\newcommand\PP{{\cal P}}
\newcommand\RR{{\cal R}}
\renewcommand\SS{{\cal S}}
\def\Sone{{{\cal S}_1}}
\def\Stwo{{{\cal S}_2}}
\newcommand\TT{{\cal T}}
\newcommand\ZZ{{\cal Z}}
\newcommand\bfa{{\mathbf a}}
\newcommand\bfA{{\mathbf A}}
\newcommand\bfb{{\mathbf b}}
\newcommand\bfB{{\mathbf B}}
\newcommand\bfc{{\mathbf c}}
\newcommand\bfC{{\mathbf C}}
\newcommand\bfd{{\mathbf d}}
\newcommand\bfD{{\mathbf D}}
\newcommand\bfe{{\mathbf e}}
\newcommand\bfE{{\mathbf E}}
\newcommand\bff{{\mathbf f}}
\newcommand\bfF{{\mathbf F}}
\newcommand\bfg{{\mathbf g}}
\newcommand\bfG{{\mathbf G}}
\newcommand\bfh{{\mathbf h}}
\newcommand\bfH{{\mathbf H}}
\newcommand\bfi{{\mathbf i}}
\newcommand\bfI{{\mathbf I}}
\newcommand\bfj{{\mathbf j}}
\newcommand\bfJ{{\mathbf J}}
\newcommand\bfk{{\mathbf k}}
\newcommand\bfK{{\mathbf K}}
\newcommand\bfl{{\mathbf l}}
\newcommand\bfL{{\mathbf L}}
\newcommand\bfm{{\mathbf m}}
\newcommand\bfM{{\mathbf M}}
\newcommand\bfn{{\mathbf n}}
\newcommand\bfN{{\mathbf N}}
\newcommand\bfo{{\mathbf o}}
\newcommand\bfO{{\mathbf O}}
\newcommand\bfp{{\mathbf p}}
\newcommand\bfP{{\mathbf P}}
\newcommand\bfq{{\mathbf q}}
\newcommand\bfQ{{\mathbf Q}}
\newcommand\bfr{{\mathbf r}}
\newcommand\bfR{{\mathbf R}}
\newcommand\bfs{{\mathbf s}}
\newcommand\bfS{{\mathbf S}}
\newcommand\bft{{\mathbf t}}
\newcommand\bfT{{\mathbf T}}
\newcommand\bfu{{\mathbf u}}
\newcommand\bfU{{\mathbf U}}
\newcommand\bfv{{\mathbf v}}
\newcommand\bfV{{\mathbf V}}
\newcommand\bfw{{\mathbf w}}
\newcommand\bfW{{\mathbf W}}
\newcommand\bfx{{\mathbf x}}
\newcommand\bfX{{\mathbf X}}
\newcommand\bfy{{\mathbf y}}
\newcommand\bfY{{\mathbf Y}}
\newcommand\bfz{{\mathbf z}}
\newcommand\bfZ{{\mathbf Z}}
\newcommand\sla{\raise.16ex\hbox{$/$}\kern-.57em}
\newcommand\Dsl{\kern.2em\raise.16ex\hbox{$/$}\kern-.77em\hbox{$D$}}
\newcommand\parsl{\sla\hbox{$\partial$}}
\newcommand\nabsl{\sla\hbox{$\nabla$}}
\newcommand\Asl{\kern.2em\raise.16ex\hbox{$/$}\kern-.77em\hbox{$A$}}
\newcommand\psl{\sla\hbox{$p$}}
\newcommand\ksl{\sla\hbox{$k$}}
\newcommand\qsl{\sla\hbox{$q$}}
\newcommand\dbar{\ensuremath{d\kern-.3em{\rule[1.3ex]{.7ex}{.06ex}}}}
\newcommand\gtwid{
 \mathrel{\raise.3ex\hbox{$>$\kern-.75em\lower1ex\hbox{$\sim$}}}}
\newcommand\ltwid{
 \mathrel{\raise.3ex\hbox{$<$\kern-.75em\lower1ex\hbox{$\sim$}}}}
\newcommand\vp{{\vec p}}
\newcommand\vx{{\vec x}}
\newcommand\su[1]{{\rm SU}(#1)}
\newcommand\so[1]{{\rm SO}(#1)}
\newcommand\you[1]{{\rm U}(#1)}
\newcommand\lefty{_{\rm L}}
\newcommand\righty{_{\rm R}}
\newcommand\lpr{_{\rm L+R}}
\newcommand\ev{{\rm\ eV}}
\newcommand\mev{{\rm\ MeV}}
\newcommand\gev{{\rm\ GeV}}
\newcommand\tev{{\rm\ TeV}}
\newcommand\kev{{\rm\ KeV}}
\newcommand\eV{{\rm\ eV}}
\newcommand\MeV{{\rm\ MeV}}
\newcommand\GeV{{\rm\ GeV}}
\newcommand\TeV{{\rm\ TeV}}
\newcommand\KeV{{\rm\ KeV}}
\newcommand\kbt{k_B T}
\newcommand\vev[1]{{\left\langle{#1}\right\rangle}}
\newcommand\bra[1]{\left\langle#1\right|}
\newcommand\ket[1]{\left|#1\right\rangle}
\newcommand\vac{\ket0}
\newcommand\sol{\ket s}
\newcommand\const{{\rm const}}
\newcommand\Tr{{\rm Tr}}
\newcommand\cm{{\rm cm}}
\newcommand\seco{{\rm sec}}
\newcommand\sech{{\rm sech}}
\newcommand\eqalignnum{\eqalignno}
\newcommand\half{{\frac 12}}
\newcommand\sq{{\vbox {\hrule height 0.6pt\hbox{\vrule width
        0.6pt\hskip 3pt
        \vbox{\vskip 6pt}\hskip 3pt \vrule width 0.6pt}
      \hrule height 0.6pt}}}
\newcommand\square{\kern1pt\vbox{\hrule height 1.2pt
    \hbox{\vrule width 1.2pt\hskip 3pt
      \vbox{\vskip 6pt}\hskip 3pt\vrule width 0.6pt}
    \hrule height 0.6pt}\kern1pt}
\newcommand\tdot[1]{\mathord{\mathop{#1}\limits^{\kern2pt\ldots}}}
\newcommand\qed{\vrule height 1.2ex width 0.5em}

%
%

\title{Path Integral Methods and Applications\thanks{Lectures given at
    Rencontres du Vietnam: VIth Vietnam School of Physics, Vung Tau,
    Vietnam, 27 December 1999 - 8 January 2000.}}
\author{Richard MacKenzie\thanks{rbmack@lps.umontreal.ca}\\
Laboratoire Ren\'e-J.-A.-L\'evesque\\
Universit\'e de Montr\'eal\\
Montr\'eal, QC H3C 3J7 Canada}
\date{\normalsize UdeM-GPP-TH-00-71}
\maketitle
\begin{abstract} These lectures are intended as an introduction to the
  technique of path integrals and their applications in physics. The
  audience is mainly first-year graduate students, and it is assumed
  that the reader has a good foundation in quantum mechanics. No prior
  exposure to path integrals is assumed, however.

  The path integral is a formulation of quantum mechanics equivalent
  to the standard formulations, offering a new way of looking at
  the subject which is, arguably, more intuitive than the usual
  approaches. Applications of path integrals are as vast as those of
  quantum mechanics itself, including the quantum mechanics of a
  single particle, statistical mechanics, condensed matter physics and
  quantum field theory.

After an introduction including a very brief historical overview of
  the subject, we derive a path integral expression for the propagator
  in quantum mechanics, including the free particle and harmonic
  oscillator as examples. We then discuss a variety of applications,
  including path integrals in multiply-connected spaces, Euclidean
  path integrals and statistical mechanics, perturbation theory in
  quantum mechanics and in quantum field theory, and instantons via
  path integrals.

For the most part, the emphasis is on explicit calculations in the
  familiar setting of quantum mechanics, with some discussion (often
  brief and schematic) of how these ideas can be applied to more
  complicated situations such as field theory.
\end{abstract}

\newpage\thispagestyle{empty}
\section{Introduction}

\subsection{Historical remarks}
We are all familiar with the standard formulations of quantum
mechanics, developed more or less concurrently by Schroe\-ding\-er,
Heisenberg and others in the 1920s, and shown to be equivalent to one
another soon thereafter.

In 1933, Dirac made the observation that the action plays a central
role in classical mechanics (he considered the Lagrangian formulation
of classical mechanics to be more fundamental than the Hamiltonian
one), but that it seemed to have no important
role in quantum mechanics as it was known at the
time. He speculated on how this situation might be rectified,
and he arrived at the conclusion that (in more modern language) the
propagator in quantum mechanics ``corresponds to'' $\exp i S/\hbar$,
where $S$ is the classical action evaluated along the classical path.

In 1948, Feynman developed Dirac's suggestion, and succeeded in
deriving a third formulation of quantum mechanics, based on the fact
that the propagator can be written as a sum over all possible
paths (not just the classical one) between the
initial and final points. Each path contributes $\exp i S/\hbar$ to
the propagator. So while Dirac considered only the classical path,
Feynman showed that all paths contribute: in a sense, the quantum
particle takes all paths, and the amplitudes for each path add
according to the usual quantum mechanical rule for combining
amplitudes. Feynman's original paper,\footnote{References
  are not cited in the
  text, but a short list of books and articles which I have found
  interesting and useful is given at the end of this article.} 
which essentially laid the foundation of the subject (and which was
rejected by Physical Review!), is an all-time classic, and is highly
recommended. (Dirac's original article is not bad, either.)

\subsection{Motivation}  
What do we learn from path integrals? As far as I am aware, path
  integrals give us no dramatic new results in the quantum mechanics
  of a single particle. Indeed, most if not all calculations in
  quantum mechaincs which can
  be done by path integrals can be done with considerably greater ease
  using the standard formulations of quantum mechanics. (It is
  probably for this reason that path integrals are often left out of
  undergraduate-level quantum mechanics courses.) So why the fuss?

As I will mention shortly, path integrals turn out to be
  considerably more useful in more complicated situations, such as
  field theory. But even if this were not the case, I believe that
  path integrals would be a very worthwhile contribution to our
  understanding of quantum mechanics. Firstly, they provide a
  physically extremely appealing and intuitive way of viewing quantum
  mechanics: anyone who can understand Young's double slit experiment
  in optics should be able to understand the underlying ideas behind
  path integrals. Secondly, the classical limit of quantum mechanics
  can be understood in a particularly clean way via path integrals.

It is in quantum field theory, both relativistic and nonrelativistic,
that path integrals (functional integrals is a more accurate term)
play a much more important role, for several reasons. They provide a
relatively easy road to quantization and to expressions for Green's
functions, which are closely related to amplitudes for physical
processes such as scattering and decays of particles.  The path
integral treatment of gauge field theories (non-abelian ones, in
particular) is very elegant: gauge fixing and ghosts appear quite
effortlessly. Also, there are a whole host of nonperturbative
phenomena such as solitons and instantons that are most easily viewed
via path integrals.  Furthermore, the close relation between
statistical mechanics and quantum mechanics, or statistical field
theory and quantum field theory, is plainly visible via path
integrals.

In these lectures, I will not have time to go into great detail into
the many useful applications of path integrals in quantum field
theory.  Rather than attempting to discuss a wide variety of
applications in field theory and condensed matter physics, and in so
doing having to skimp on the ABCs of the subject, I have chosen to
spend perhaps more time and effort than absolutely necessary showing
path integrals in action (pardon the pun) in quantum mechanics. The
main emphasis will be on quantum mechanical problems which are not
necessarily interesting and useful
in and of themselves, but whose principal
value is that they resemble the calculation of similar objects in the
more complex setting of quantum field theory, where explicit
calculations would be much harder. Thus I hope to illustrate the main
points, and some technical complications and hangups which arise, in
relatively familiar situations that should be regarded as toy models
analogous to some interesting contexts in field theory.

\subsection{Outline}
The outline of the lectures is as follows. In the next section I will
begin with an introduction to path integrals in quantum mechanics,
including some explicit examples such as the free particle and the
harmonic oscillator. In Section 3, I will give a ``derivation'' of
classical mechanics from quantum mechanics. In Section 4, I will
discuss some applications of path integrals that are perhaps not so
well-known, but nonetheless very amusing, namely, the case where the
configuration space is not simply connected. (In spite of the fancy
terminology, no prior knowledge of high-powered mathematics such as
topology is assumed.) Specifically, I will apply the method to the
Aharonov-Bohm effect, quantum statistics and anyons, and monopoles and
charge quantization, where path integrals provide a beautifully
intuitive approach. In Section 5, I will explain how one can approach
statistical mechanics via path integrals. Next, I will discuss
perturbation theory in quantum mechanics, where the technique used is
(to put it mildly) rather cumbersome, but nonetheless
illustrative for applications in the remaining
sections. In Section 7, I will discuss Green's functions (vacuum
expectation values of time-ordered products) in quantum mechanics
(where, to my knowledge, they are not particularly useful), and will
construct the generating functional for these objects. This groundwork
will be put to good use in the following section, where the generating
functional for Green's functions in field theory (which {\em are}
useful!) will be elucidated. In Section 9, I will discuss instantons
in quantum mechanics, and will at least pay lip service to important
applications in field theory. I will finish with a summary and a list of
embarrassing omissions.

\vspace{1cm}
I will conclude with a few apologies.
First, an educated reader might get the impression that the outline
given above contains for the most part standard material. S/he is
likely correct: the only original content to these lectures is the
errors.\footnote{Even this joke is borrowed from somewhere, though
I can't think of where.}

Second, I have made no great
effort to give complete references (I know my limitations); 
at the end of this article I have listed some papers and
books from which I have learned the subject. Some are books or
articles wholly devoted to path integrals; the majority are books for
which path integrals form only a small (but interesting!)
part. The list is
hopelessly incomplete; in particular, virtually any quantum field
theory book from the last decade or so has a discussion of path
integrals in it.

Third, the subject of path integrals can be a rather delicate one for
the mathematical purist. I am not one, and I have neither the interest
nor the expertise to go into detail about whether or not the path
integral exists, in a strict sense. My approach is rather pragmatic:
it works, so let's use it!

\newpage\thispagestyle{empty}
\section{Path Integrals in Quantum Mechanics}

\subsection{General discussion}
Consider a particle moving in one dimension, the Hamiltonian being of
the usual form:
\[ H={p^2\over2m}+V(q).\]
The fundamental question in the path integral (PI) formulation of
quantum mechanics is: If the particle is at a position $q$ at
time $t=0$, what is the probability amplitude that it will be at some
other position $q'$ at a later time $t=T$?

It is easy to get a formal expression for
this amplitude in the usual Schroedinger
formulation of quantum mechanics. Let us introduce the
eigenstates of the position operator $\hat q$, which form a complete,
orthonormal set:
\[ \hat q\ket q=q\ket q,\qquad \bra{q'}q\rangle=\delta(q'-q),
\qquad \int dq\ket q\bra q=1.\]
(When there is the possibility of an ambiguity, operators will be 
written with a ``hat''; otherwise the hat will be dropped.)
Then the initial state is $\ket{\psi(0)}=\ket q$. Letting the state
evolve in time and projecting on the state $\ket{q'}$, we get for the
amplitude $A$,
\beq A=\bra{q'}\psi(T)\rangle\equiv K(q',T;q,0)
=\bra{q'}e^{-iHT}\ket q.\label{ampl} \eeq
(Except where noted otherwise, $\hbar$ will be set to 1.)
This object, for obvious reasons, is known as the propagator from the
initial spacetime point $(q,0)$ to the final point $(q',T)$. Clearly,
the propagator is independent of the origin of time:
$K(q',T+t;q,t)=K(q',T;q,0)$.

We will derive an expression for this amplitude in the form of a
summation (integral, really) over all possible paths between the
initial and final points. In so doing, we derive the PI
from quantum mechanics. Historically, Feynman came up with the PI
differently, and showed its equivalence to the usual formulations
of quantum mechanics.

Let us separate the time evolution in the above amplitude into two
smaller time evolutions, writing $e^{-iHT}=e^{-iH(T-t_1)}e^{-iHt_1}$.
The amplitude becomes
\[ A=\bra{q'}e^{-iH(T-t_1)}e^{-iHt_1}\ket q.\]
Inserting a factor 1 in the form of a sum over the position
eigenstates gives
\bea A&=&\bra{q'}e^{-iH(T-t_1)}
\underbrace{\int dq_1\ket{q_1}\bra{q_1}}_{=1}
e^{-iHt_1}\ket q\nonumber\\
&=&\int dq_1\,K(q',T;q_1,t_1)K(q_1,t_1;q,0).\label{kk}\eea
This formula is none other than an expression of the quantum mechanical
rule for combining amplitudes: if a process can occur a number of
ways, the amplitudes for each of these ways add. A
particle, in propagating from $q$ to $q'$, must be {\em somewhere} at an
intermediate time $t_1$; labelling that intermediate position $q_1$,
we compute the amplitude for propagation via the point $q_1$ [this is
the product of the two propagators in \eqref{kk}] and integrate over all
possible intermediate positions. This result is reminiscent of Young's
double slit experiment, where the amplitudes for passing through each
of the two slits combine and interfere. We will look at the
double-slit experiment in more detail when we discuss the Aharonov-Bohm
effect in Section 4.

We can repeat the division of the time interval $T$; let us divide it
up into a large number $N$ of time intervals of
duration $\delta=T/N$. Then we can
write for the propagator
\[ A=\bra{q'}\left(e^{-iH\delta}\right)^N\ket q
=\bra{q'}\underbrace{e^{-iH\delta}e^{-iH\delta}\cdots 
e^{-iH\delta}}_{N\ \mathrm{times}}\ket q.\]
We can again insert a complete set of states between each exponential,
yielding
\bea
A&=&\bra{q'}e^{-iH\delta}\int dq_{N-1}\ket{q_{N-1}}\bra{q_{N-1}}
e^{-iH\delta}\int dq_{N-2}\ket{q_{N-2}}\bra{q_{N-2}}\cdots\nonumber\\
&&\qquad\cdots\int dq_{2}\ket{q_{2}}\bra{q_{2}}e^{-iH\delta}
\int dq_{1}\ket{q_{1}}\bra{q_{1}}e^{-iH\delta}\ket{q}\nonumber\\
&=&\int dq_1\cdots dq_{N-1}\bra{q'}e^{-iH\delta}
\ket{q_{N-1}}\bra{q_{N-1}}e^{-iH\delta}\ket{q_{N-2}}\cdots\nonumber\\
&&\qquad\cdots\bra{q_{1}}e^{-iH\delta}\ket{q}\nonumber\\
&\equiv&\int dq_1\cdots dq_{N-1}K_{q_N,q_{N-1}}K_{q_{N-1},q_{N-2}}\cdots
K_{q_{2},q_{1}}K_{q_{1},q_{0}},\label{kkkk}
\eea
where we have defined $q_0=q$, $q_N=q'$. (Note that these initial and
final positions are {\em not} integrated over.)
This expression says that the amplitude is the integral of the
amplitude of all $N$-legged paths, as illustrated in Figure 1.
\begin{figure}[hb]
\epsfysize=5cm
\centerline{\epsfbox{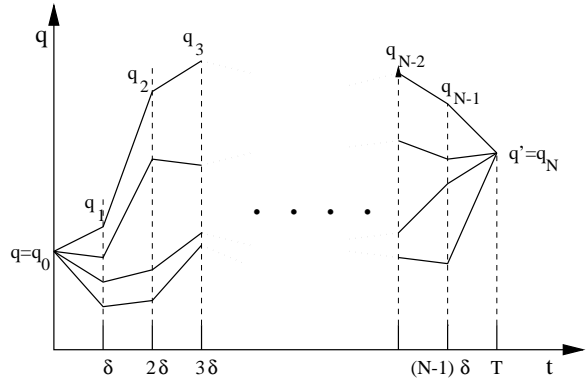}}
\caption{Amplitude as a sum over all $N$-legged paths.}
\end{figure}

Apart from mathematical details concerning the limit when
$N\to\infty$, this is clearly going to become a sum over all
possible paths of the amplitude for each path:
\[ A=\sum_{\mathrm{paths}}A_{\mathrm{path}},\]
where
\[ \sum_{\mathrm{paths}}=\int dq_1\cdots dq_{N-1},\qquad
A_{\mathrm{path}}=K_{q_N,q_{N-1}}K_{q_{N-1},q_{N-2}}\cdots
K_{q_{2},q_{1}}K_{q_{1},q_{0}}.\]
Let us look at this last expression in detail.

The propagator for one sub-interval is 
$K_{q_{j+1},q_j}=\bra{q_{j+1}}e^{-iH\delta}\ket{q_j}$. We can expand
the exponential, since $\delta$ is small:
\bea 
K_{q_{j+1},q_j}&=&\bra{q_{j+1}}\left(1-iH\delta-\half
  H^2\delta^2+\cdots\right)\ket{q_j}\nonumber\\
&=&\bra{q_{j+1}}q_j\rangle-i\delta\bra{q_{j+1}}H\ket{q_j}
  +o(\delta^2).
\label{kqq1}
\eea
The first term is a delta function, which we can write\footnote{Please
do not confuse the delta function with the time interval, $\delta$.}
\beq
\bra{q_{j+1}}q_j\rangle=\delta(q_{j+1}-q_j)
=\int{dp_j\over2\pi}e^{ip_j(q_{j+1}-q_j)}.
\label{kqq1a}
\eeq
In the second term of \eqref{kqq1}, 
we can insert a factor 1 in the form of an integral over momentum
eigenstates between $H$ and
$\ket{q_j}$; this gives
\bea &&-i\delta\bra{q_{j+1}}\left({{\hat p}^2\over2m}+V(\hat q)\right)
\int{dp_j\over2\pi}\ket{p_j}\bra{p_j}q_j\rangle\nonumber\\
&&\quad=-i\delta\int{dp_j\over2\pi}
\left({{p_j}^2\over2m}+V(q_{j+1})\right)\bra{q_{j+1}}p_j\rangle
\bra{p_j}q_j\rangle\nonumber\\
&&\quad=-i\delta\int{dp_j\over2\pi}
\left({{p_j}^2\over2m}+V(q_{j+1})\right)e^{ip_j(q_{j+1}-q_j)},
\label{kqq2}\eea
using $\bra{q}p\rangle=\exp ipq$.
In the first line, we view the operator $\hat p$ as operating to the
right, while $V(\hat q)$ operates to the left. 

The expression
\eqref{kqq2} is asymmetric between $q_j$ and $q_{j+1}$; the origin of
this is our choice of putting the factor 1 to the right of $H$ in the
second term of \eqref{kqq1}. Had we put it to the left
instead, we would have
obtained $V(q_{j})$ in \eqref{kqq2}. To not play favourites, we
should choose some sort of average of these two. In what follows I
will simply write $V({\bar q}_j)$ where
${\bar q}_j=\half(q_j+q_{j+1})$. (The exact choice does not matter in
the continuum limit, which we will take eventually; the above is a
common choice.) Combining \eqref{kqq1a} and \eqref{kqq2}, 
the sub-interval propagator is
\bea
K_{q_{j+1},q_j}&=&\int{dp_j\over2\pi}e^{ip_j(q_{j+1}-q_j)}
\left(1-i\delta\left({{p_j}^2\over2m}+V({\bar q}_j)\right)
+o(\delta^2)\right)\nonumber\\
&=&\int{dp_j\over2\pi}e^{ip_j(q_{j+1}-q_j)}
e^{-i\delta H(p_j,{\bar q}_j)}(1+o(\delta^2)).\label{kqq3}\eea
There are $N$ such factors in the amplitude. Combining them, and
writing ${\dot q}_j=(q_{j+1}-q_j)/\delta$, we get
\beq 
A_{\mathrm{path}}=\int\prod_{j=0}^{N-1}{dp_j\over2\pi}
\exp i\delta\sum_{j=0}^{N-1}(p_j 
      {\dot q}_j-H(p_j,{\bar q}_j)),
\label{apath} 
\eeq
where we have neglected a multiplicative
factor of the form $(1+o(\delta^2))^N$,
which will tend toward one in the continuum limit.
Then the propagator becomes
\bea
K&=&\int dq_1\cdots dq_{N-1}A_{\mathrm{path}}\nonumber\\
&=&\int\prod_{j=1}^{N-1}dq_j\int\prod_{j=0}^{N-1}{dp_j\over2\pi}
\exp i\delta\sum_{j=0}^{N-1}(p_j 
      {\dot q}_j-H(p_j,{\bar q}_j)).
\label{pi1}
\eea
Note that there is one momentum integral for each interval ($N$
total), while there is one position integral for each intermediate
position ($N-1$ total).

If $N\to\infty$, this approximates an integral over all functions
$p(t)$, $q(t)$. We adopt the following notation:
\beq 
\fbox{$\displaystyle
K\equiv\int \DD p(t)\DD q(t)
\exp i\int_0^T dt\left(p\dot q-H(p,q)\right).
$}
\label{phase}
\eeq
This result is known as the {\em phase-space path integral}.
The integral is viewed as over all functions $p(t)$ and over all
functions $q(t)$ where $q(0)=q$, $q(T)=q'$. But to actually perform an
explicit calculation, \eqref{phase} should be viewed as a shorthand
notation for the more ponderous expression \eqref{pi1}, in the limit
$N\to\infty$.

If, as is often the case (and as we have assumed in deriving the above
expression), the Hamiltonian is of the standard form, namely
$H=p^2/2m+V(q)$, we can actually carry out the momentum
integrals in
\eqref{pi1}. We can rewrite this expression as
\[ K=\int\prod_{j=1}^{N-1}dq_j 
\exp-i\delta\sum_{j=0}^{N-1} V({\bar q}_j)
\int\prod_{j=0}^{N-1}{dp_j\over2\pi}
\exp i\delta\sum_{j=0}^{N-1}
\left(p_j {\dot q}_j-{p_j}^2/2m\right). \]
The $p$ integrals are all Gaussian, and they are uncoupled. One such
integral is
\[ \int{dp\over2\pi}e^{i\delta(p{\dot q}-p^2/2m)}=
\sqrt{m\over2\pi i\delta}e^{i\delta m{\dot q}^2/2}.\]
(The careful reader may be worried about the convergence of this
integral; if so, a factor $\exp-\eps p^2$ can be introduced and the
limit $\eps\to0$ taken at the end.)

The propagator becomes
\bea
K&=&\int\prod_{j=1}^{N-1}dq_j
\exp-i\delta\sum_{j=0}^{N-1} V({\bar q}_j)
\prod_{j=0}^{N-1}\left(\sqrt{m\over2\pi i\delta}
\exp i\delta{m{\dot q}_j^2\over2}\right)\nonumber\\
&=&\left({m\over2\pi i\delta}\right)^{N/2}
\int\prod_{j=1}^{N-1}dq_j
\exp i\delta\sum_{j=0}^{N-1}
\left({m{\dot q}_j^2\over2}-V({\bar q}_j)\right).
\label{pi2}
\eea
The argument of the exponential is a discrete approximation of the
action of a path passing through the points
$q_0=q,q_1,\cdots,q_{N-1},q_N=q'$. As above, 
we can write this in the more
compact form
\beq
\fbox{$\displaystyle
K=\int\DD q(t) e^{i S[q(t)]}.
$}
\label{config}
\eeq
This is our final result, and is known as the {\em
configuration space path
integral}. Again, \eqref{config} should be viewed as a notation for the
more precise expression \eqref{pi2}, as $N\to\infty$.

\subsection{Examples}
To solidify the notions above, let us consider a few explicit
examples. As a first example, we will
compute the free particle propagator first using ordinary
quantum mechanics and then via the PI. We will then mention
some generalizations which can be done in a similar manner.

\subsubsection{Free particle} 
Let us compute the propagator $K(q',T;q,0)$ for a free particle,
described by the Hamiltonian $H=p^2/2m$. The propagator can be
computed straightforwardly using ordinary quantum mechanics. To this
end, we write
\bea
K&=&\bra{q'}e^{-iHT}\ket q\nonumber\\
&=&\bra{q'}e^{-iT{\hat p}^2/2m}\int{dp\over2\pi}
\ket{p}\bra{p}q\rangle\nonumber\\
&=&\int{dp\over2\pi}e^{-iT{p}^2/2m}\bra{q'}p\rangle
\bra{p}q\rangle\nonumber\\
&=&\int{dp\over2\pi}e^{-iT({p}^2/2m)+i(q'-q)p}.
\label{free1}
\eea
The integral is Gaussian; we obtain
\beq
K=\left({m\over2\pi iT}\right)^{1/2}e^{im(q'-q)^2/2T}.
\label{free2}
\eeq

Let us now see how the same result can be attained using PIs.
The configuration space PI \eqref{config} is
\bea
K&=&\lim_{N\to\infty}\left({m\over2\pi i\delta}\right)^{N/2}
\int\prod_{j=1}^{N-1}dq_j\,\exp i{m\delta\over2}\sum_{j=0}^{N-1}
\left({q_{j+1}-q_j\over\delta}\right)^2\nonumber\\
&=&\lim_{N\to\infty}\left({m\over2\pi i\delta}\right)^{N/2}
\int\prod_{j=1}^{N-1}dq_j\,\exp i{m\over2\delta}
\Big[(q_{N}-q_{N-1})^2+(q_{N-1}-q_{N-2})^2+\cdots\nonumber\\
&&\qquad\qquad\qquad\qquad\qquad\qquad
+(q_{2}-q_{1})^2+(q_{1}-q_{0})^2\Big],\nonumber
\eea
where $q_0=q$ and $q_N=q'$ are the initial and final points.
The integrals are Gaussian, and can be evaluated exactly, although the
fact that they are coupled complicates matters significantly. 
The result is
\bea
K&=&\lim_{N\to\infty}\left({m\over2\pi i\delta}\right)^{N/2}
{1\over\sqrt N}\left({2\pi i\delta\over m}\right)^{(N-1)/2}
e^{im(q'-q)^2/2N\delta}\nonumber\\
&=&\lim_{N\to\infty}\left({m\over2\pi iN\delta}\right)^{1/2}
e^{im(q'-q)^2/2N\delta}.\nonumber
\eea
But $N\delta$ is the total time interval $T$, resulting in
\[
K=\left({m\over2\pi iT}\right)^{1/2}
e^{im(q'-q)^2/2T},
\]
in agreement with \eqref{free2}.

A couple of remarks are in order. First, we can write the argument of
the exponential as $T\cdot\half m((q'-q)/T)^2$, which is just the
action $S[q_c]$
for a particle moving along the classical path (a straight line
in this case) between the initial and final points. 

Secondly, we can
restore the factors of $\hbar$ if we want, by ensuring correct
dimensions. The argument of the exponential is the action, so in order
to make it a pure number we must divide by $\hbar$; furthermore, the
propagator has the dimension of the inner product of two position
eigenstates, which is inverse length; in order that the coefficient
have this dimension we must multiply by $\hbar^{-1/2}$. The final
result is
\beq 
\fbox{$\displaystyle
K=\left({m\over2\pi i\hbar T}\right)^{1/2}e^{iS[q_c]/\hbar}.
$}
\label{freepart}
\eeq
This result typifies a couple of important features of calculations in
this subject, which we will see repeatedly in these lectures.
First,
the propagator separates into two factors, one of which is
the phase $\exp iS[q_c]/\hbar$. 
Second, calculations in the PI formalism are typically quite a
bit more lengthy than using standard techniques
of quantum mechanics.

\subsubsection{Harmonic oscillator}
As a second example of the computation of a PI, let us
compute the propagator for the harmonic oscillator using this method.
(In fact, we will not do the entire computation, but we will do enough
to illustrate a trick or two which will be useful later on.)

Let us start with the
somewhat-formal version of the configuration-space PI,
\eqref{config}:
\[ K(q',T;q,0)=\int\DD q(t) e^{i S[q(t)]}.
\]
For the harmonic oscillator,
\[ S[q(t)]=\int_0^T dt\left(\half m{\dot q}^2 
-\half m\om^2q^2 \right).
\]
The paths over which the integral is to
be performed go from $q(0)=q$ to
$q(T)=q'$. To do this PI, suppose we know the solution of the
classical problem, $q_c(t)$:
\[{\ddot q}_c+\om^2q_c=0,\qquad q_c(0)=q,\qquad q_c(T)=q'.\]
We can write $q(t)=q_c(t)+y(t)$, and perform a change of variables in
the PI to $y(t)$, since integrating over all deviations from the
classical path is equivalent to integrating over all possible
paths. Since at each time $q$ and $y$ differ by a constant,
the Jacobian of the transformation is 1. Furthermore, since $q_c$
obeys the correct boundary conditions, the paths $y(t)$ over which we
integrate go from $y(0)=0$ to $y(T)=0$.
The action for the path $q_c(t)+y(t)$ can be
written as a power series in $y$:
\beano
S[q_c(t)+y(t)]&=&\int_0^T dt\left(\half m{{\dot q}_c}^2
-\half m\om^2{q_c}^2\right)
+\underbrace{(\mbox{linear in $y$})}_{=0}
+\int_0^T dt \left(\half m {\dot y}^2
-\half m\om^2 y^2\right).
\eeano
The term linear in $y$ vanishes by construction:
$q_c$, being the classical path, is that path for which the action is
stationary! So we may write
$S[q_c(t)+y(t)]=S[q_c(t)]+S[y(t)]$. We substitute this into
\eqref{config}, yielding
\beq
K(q',T;q,0)=e^{i S[q_c(t)]}\int\DD y(t) e^{i S[y(t)]}.
\label{harmosc1}\eeq
As mentioned above, the paths $y(t)$ over which we integrate
go from $y(0)=0$ to $y(T)=0$: the only appearance of the
initial and final positions is in the classical path, \ie in the
classical action. Once again, the PI separates 
into two factors. The first is written in terms of
the action of the classical path, and the second is a PI
over deviations from this classical path. The second factor is
independent of the initial and final points.

This separation into a factor depending on the
action of the classical path and a second one, a PI which
is independent of the details of the classical
path, is a recurring theme, and an important one.
Indeed, it is often the first
factor which contains most of the useful information contained
in the propagator,
and it can be deduced without even
performing a PI. It can be said that much of the work in
the game of path integrals consists
in avoiding having to actually compute one!

As for the evaluation of \eqref{harmosc1}, a number of fairly standard
techniques are available. One can calculate
the PI directly in position space,
as was done above for the harmonic oscillator (see Schulman,
chap.~6). Alternatively, one can compute it in Fourier space
(writing $y(t)=\sum_k a_k\sin(k\pi t/T)$ and integrating over the
coefficients $\{a_k\}$). This latter approach is outlined in Feynman
and Hibbs, Section 3.11. The result is
\beq
K(q',T;q,0)=\left({m\om\over2\pi i\sin\om T}\right)^{1/2}
e^{i S[q_c(t)]}.
\label{harmosc2}\eeq

The classical action can be evaluated straightforwardly (note that this
is not a PI problem, nor even a quantum mechanics problem!);
the result is
\[ S[q_c(t)]={m\om\over2\sin\om T}
\left(({q'}^2+q^2)\cos\om T-2q'q\right).\]

We close this section with two remarks. First,
the PI for any quadratic action
can be evaluated exactly, essentially since such a PI consists of
Gaussian integrals; the general result is given in Schulman, Chapter 6.
In Section 6, we will evaluate (to the same degree of completeness as
the harmonic oscillator above) the PI for a forced harmonic
oscillator, which will prove to be a very useful tool for computing a
variety of quantities of physical interest.

Second, the following fact is not difficult to prove, and will be used
below (Section 4.2.). $K(q',T;q,0)$ (whether computed via PIs or not)
is the amplitude to propagate from
one point to another in a given time interval. But this is the
response to the following question: If a particle is initially at
position $q$, what is its wave function after the elapse of a time
$T$? Thus, if we consider $K$ as a
function of the final position and time, it is none other than the
wave function for a particle with a specific initial condition. As
such, the propagator satisfies the Schroedinger equation at its final
point.

\newpage\thispagestyle{empty}
\section[Classical Limit]
{The Classical Limit: ``Derivation'' of the Principle of Least Action}

Since the example calculations performed above are somewhat dry and
mathematical, it is worth backing up a bit and staring at the
expression for the configuration space PI, \eqref{config}:
\[ K=\int\DD q(t) e^{i S[q(t)]/\hbar}.
\]
This innocent-looking expression tells us something which is at first
glance unbelievable, and at second glance {\em really} unbelievable.
The first-glance observation is that a particle, in going from one
position to another, takes all possible paths between these two
positions. This is, if not actually unbelievable, at the very least
least counter-intuitive, but we could argue away much of what makes us
feel uneasy if we could convince ourselves that while all paths
contribute, the classical path is the dominant one. 

However, the second-glance observation is not reassuring: if we
compare the contribution of the classical path (whose action is
$S[q_c]$) with that of some other, arbitrarily wild, path (whose
action is $S[q_w]$), we find that the first is $\exp iS[q_c]$ while
the second is $\exp iS[q_w]$. They are both complex numbers of {\em
unit magnitude}: each path taken in isolation is {\em equally
important}. The classical path is no more important than any
arbitrarily complicated path!

How are we to reconcile this {\em really} unbelievable conclusion with
the fact that a ball thrown in the air has a more-or-less parabolic
motion?

The key, not surprisingly, is in how different paths interfere with
one another, and by considering the case where the rough scale of
classical action of the problem is much bigger than the quantum of
action, $\hbar$, we will see the emergence of the Principle of Least
Action.

Consider two neighbouring paths $q(t)$ and $q'(t)$
which contribute to the PI
(Figure 2). Let $q'(t)=q(t)+\eta(t)$, with $\eta(t)$ small. Then we
can write the action as a functional Taylor expansion about the
classical path:\footnote{The reader unfamiliar with manipulation of
  functionals need not despair; the only rule needed beyond
  standard calculus is the functional derivative: $\delta q(t)/\delta
  q(t')=\delta(t-t')$, where the last $\delta$ is the Dirac delta
  function.}
\[ S[q']=S[q+\eta]=S[q]+\int dt\,\eta(t){\delta S[q]\over\delta q(t)}
+o(\eta^2). \]
The two paths contribute $\exp iS[q]/\hbar$ and $\exp iS[q']/\hbar$ to
the PI; the combined contribution is
\[ 
A\simeq e^{iS[q]/\hbar}\left(1+\exp {i\over\hbar}
\int dt\,\eta(t){\delta S[q]\over\delta q(t)}\right),
\]
where we have neglected corrections of order $\eta^2$. We see that the
difference in phase between the two paths, which determines the
interference between the two contributions, is 
$\hbar^{-1}\int dt\,\eta(t)\delta S[q]/\delta q(t)$.
\begin{figure}[ht]
\epsfysize=5cm
\centerline{\epsfbox{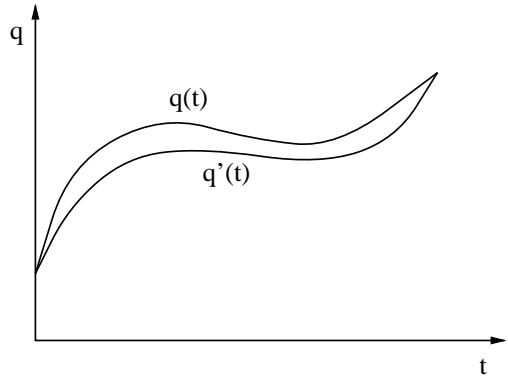}}
\caption{Two neighbouring paths.}
\end{figure}

We see that the smaller the value of $\hbar$, the larger the phase
difference between two given paths. So even if the paths are {\em very}
close together, so that the difference in actions is extremely small,
for sufficiently small $\hbar$ the phase difference will still be
large, and on average destructive interference occurs.

However, this argument must be rethought for one exceptional path: that
which extremizes the action, \ie the classical path, $q_c(t)$. 
For this path, $S[q_c+\eta]=S[q_c]+o(\eta^2)$. Thus the classical path
and a very close neighbour will have actions which differ by much less
than two randomly-chosen but equally close paths (Figure 3).
\begin{figure}[hb]
\epsfysize=5cm
\centerline{\epsfbox{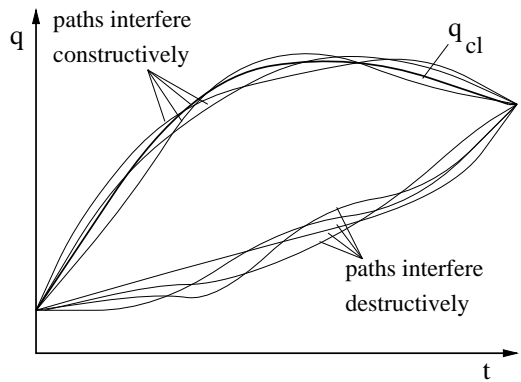}}
\caption{Paths near the classical path interfere constructively.}
\end{figure}
This means that for fixed closeness of two paths (I leave it as an
exercise to make this precise!) and for fixed $\hbar$, paths near the
classical path will on average interfere constructively (small phase
difference) whereas for random paths the interference will be on
average destructive.

Thus heuristically, we conclude that if the problem is classical
(action $\gg\hbar$), the most important contribution to the PI comes
from the region around the path which extremizes the PI. In other
words, the particle's motion is governed by the principle that the
action is stationary. This, of course, is none other than the
Principle of Least Action from which the Euler-Lagrange equations of
classical mechanics are derived.

\newpage\thispagestyle{empty}
\section[Topology and Path Integrals]
{Topology and Path Integrals in Quantum Mechanics: Three Applications}

In path integrals, if the configuration space has holes in it such
that two paths between the same initial and final point are
not necessarily deformable into one another, interesting effects can
arise. This property of the configuration space goes by the following
catchy name: non-sim\-ply-con\-nec\-ted\-ness. We will study three
such situations: the Aharonov-Bohm effect, particle statistics, and
magnetic monopoles and the quantization of electric charge.

\subsection{Aharonov-Bohm effect}
The Aharonov-Bohm effect is one of the most dramatic illustrations of
a purely quantum effect: the influence of the electromagnetic
potential on particle motion even if the particle is {\em perfectly}
shielded from any electric or magnetic fields. While classically the
effect of electric and magnetic fields can be understood purely in
terms of the forces these fields create on particles, Aharonov and
Bohm devised an ingenious thought-experiment (which has since been
realized in the laboratory) showing that this is no longer true in
quantum mechanics. Their effect is best illustrated by a refinement of
Young's double-slit experiment, where particles passing through a
barrier with two slits in it produce an interference
pattern on a screen further downstream.
Aharonov and Bohm proposed such an experiment performed with
charged particles, with an added twist provided
by a magnetic flux from which
the particles are perfectly shielded passing between the two slits.
\begin{figure}[hb]
\epsfysize=7cm
\centerline{\epsfbox{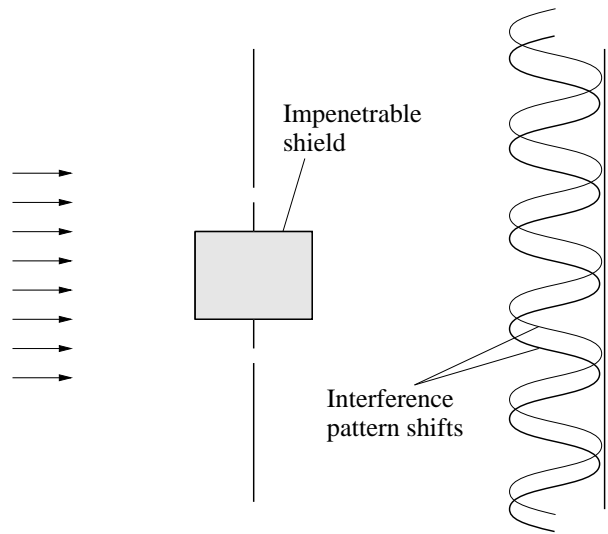}}
\caption[Aharonov-Bohm effect.]{Aharonov-Bohm effect. Magnetic flux
is confined within the shaded area; particles are excluded from
this area by a perfect shield.}
\end{figure}
If we perform the experiment first with no magnetic flux and then with
a nonzero and arbitrary flux passing through the shielded region, the
interference pattern will change, {\em in spite of the fact that the
  particles are perfectly shielded from the magnetic field and feel no
  electric or magnetic force whatsoever}. Classically we can say: no
force, no effect. Not so in quantum mechanics.
PIs provide a very attractive way of
understanding this effect.

Consider first two representative
paths $\bfq_1(t)$ and $\bfq_2(t)$ (in two dimensions)
passing through slits 1 and 2, respectively, and which arrive at
the same spot on the screen (Figure 5). 
Before turning on the magnetic field,
let us suppose that the actions for these paths are $S[\bfq_1]$ and
$S[\bfq_2]$. Then the interference of the amplitudes is determined by
\[ 
e^{iS[\bfq_1]/\hbar}+e^{iS[\bfq_2]/\hbar}
=e^{iS[\bfq_1]/\hbar}
\left(1+e^{i(S[\bfq_2]-S[\bfq_1])/\hbar}\right).
\]
The relative phase is $\phi_{12}\equiv(S[\bfq_2]-S[\bfq_1])/\hbar$. 
Thus these two paths
interfere constructively if $\phi_{12}=2n\pi$,
destructively if $\phi_{12}=(2n+1)\pi$, and in general there is partial
cancellation between the two contributions.
\begin{figure}[hb]
\epsfysize=5cm
\centerline{\epsfbox{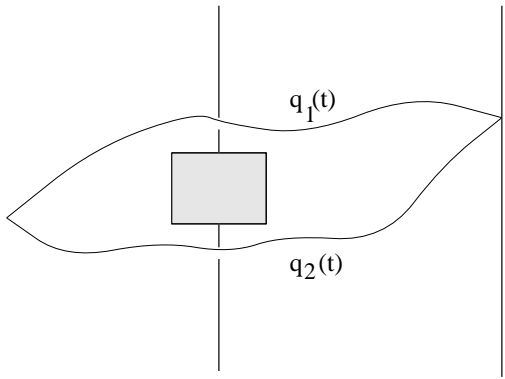}}
\caption{Two representative paths contributing to the amplitude for a
  given point on the screen.}
\end{figure}

How is this result affected if we add a magnetic field, $\bfB$? 
We can describe
this field by a vector potential, 
writing $\bfB=\nabla\times\bfA$. This
affects the particle's motion by the following change in the
Lagrangian:
\[ L(\dot \bfq,\bfq)\to L'(\dot \bfq,\bfq)=L(\dot \bfq,\bfq)-{e\over
  c}\bfv\cdot\bfA(\bfq). \]
Thus the action changes by
\[ -{e\over c}\int dt\,\bfv\cdot\bfA(\bfq)=
-{e\over c}\int dt\,{d\bfq(t)\over dt}\cdot\bfA(\bfq(t)). \]
This integral is $\int d\bfq\cdot\bfA(\bfq)$, the line integral of
$\bfA$ along the path taken by the particle. So including the effect
of the magnetic field, the action of the first path is
\[ S'[\bfq_1]=S[\bfq_1]-{e\over c}
\int_{\bfq_1(t)} d\bfq\cdot\bfA(\bfq), 
\]
and similarly for the second path.

Let us now look at the interference between the two paths, including
the magnetic field.
\bea
e^{iS'[\bfq_1]/\hbar}+e^{iS'[\bfq_2]/\hbar}&=&
e^{iS'[\bfq_1]/\hbar}\left(1+e^{i(S'[\bfq_2]-S'[\bfq_1])/\hbar}\right)
\nonumber\\
&=&e^{iS'[\bfq_1]/\hbar}\left(1+e^{i\phi'_{12}}\right),
\label{ab1}\eea
where the new relative phase is
\beq
\phi'_{12}=\phi_{12}-{e\over\hbar c}
\left(\int_{\bfq_2(t)} d\bfq\cdot\bfA(\bfq)-
\int_{\bfq_1(t)} d\bfq\cdot\bfA(\bfq)\right).
\label{ab2}
\eeq
But the difference in line integrals in \eqref{ab2} is a contour
integral:
\[
\int_{\bfq_2(t)} d\bfq\cdot\bfA(\bfq)-
\int_{\bfq_1(t)} d\bfq\cdot\bfA(\bfq)=\oint
d\bfq\cdot\bfA(\bfq)=\Phi,
\]
$\Phi$ being the flux inside the closed loop bounded by
the two paths. So we can write
\[ \phi'_{12}=\phi_{12}-{e\Phi\over\hbar c}.
\]

It is important to note that the change of relative phase due
to the magnetic field is independent of the details
of the two paths, as long as each passes through
the corresponding slit. 
This means that the PI expression for the amplitude for the particle
to reach a given point on the screen is affected by the magnetic field
in a particularly clean way. Before the magnetic field is
turned on, we may write $A=A_1+A_2$, where
\[ A_1=\int_{\mathrm{slit\ 1}}\DD \bfq\,e^{iS[\bfq]/\hbar},
\]
and similarly for $A_2$. Including the magnetic field,
\[ A_1'=\int_{\mathrm{slit\ 1}}\DD \bfq\,
e^{i(S[\bfq]-(e/c)\int d\bfq\cdot\bfA)/\hbar}
=e^{{-ie\int_1 d\bfq\cdot\bfA}/\hbar c} A_1,
\]
where we have pulled the line integral out of the PI since it is the
same for all paths passing through slit 1 arriving at the point on the
screen under consideration. So the amplitude is
\beano
A&=&e^{-ie\int_1 d\bfq\cdot\bfA/\hbar c} A_1
+e^{-ie\int_2 d\bfq\cdot\bfA/\hbar c} A_2\\
&=&e^{-ie\int_1 d\bfq\cdot\bfA/\hbar c}
\left(A_1+e^{-ie\oint d\bfq\cdot\bfA/\hbar c}
 A_2\right)\\
&=&e^{-ie\int_1 d\bfq\cdot\bfA/\hbar c}
\left(A_1+e^{-ie\Phi/\hbar c} A_2\right).
\eeano
The overall phase is irrelevant, and the interference pattern is
influenced directly by the phase $e\Phi/\hbar c$.
If we vary this phase continuously (by varying
the magnetic flux), we can detect a shift in the interference
pattern. For example, if $e\Phi/\hbar c=\pi$, then a spot on the
screen which formerly corresponded to constructive interference will
now be destructive, and vice-versa.

Since the interference is dependent only on the phase difference mod
$2\pi$, as we vary the flux we get a shift of the interference pattern
which is periodic, repeating itself when $e\Phi/\hbar c$ changes by an
integer times $2\pi$.

\subsection{Particle Statistics}

The path integral can be used to see that particles in three
dimensions
must obey either
Fermi or Bose statistics, whereas particles in two dimensions can
have intermediate (or fractional) statistics. Consider a system of two
identical particles; suppose that there is a short range, infinitely
strong repulsive force between the two. We might ask the following
question: if at $t=0$ the particles are at $\bfq_1$ and $\bfq_2$, what
is the amplitude that the particles will be at $\bfq_1'$ and 
$\bfq_2'$ at some later time $T$? We will first examine this question
in three dimensions, and then in two dimensions.

\subsubsection{Three dimensions}

According to the PI description of the problem, this amplitude is
\[ A=\sum_{\mathrm{paths}}e^{iS[\bfq_1(t),\bfq_2(t)]},
\]
where we sum over all two-particle paths going from $\bfq_1,\bfq_2$ to
$\bfq_1',\bfq_2'$.

However there is an important subtlety at play: if the particles are
identical, then there are (in three dimensions!) two classes of paths
(Figure 6).
\begin{figure}[hb]
\epsfysize=4cm
\centerline{\epsfbox{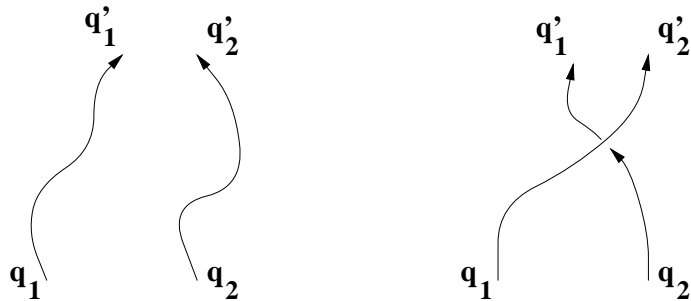}}
\caption{Two classes of paths.}
\end{figure}

Even though the second path involves an exchange of particles, the
final configuration is the same due to the indistinguishability of the
particles.

It is more economical to describe this situation in terms of the
centre-of-mass position $\bfQ=(\bfq_1+\bfq_2)/2$ and the relative
position $\bfq=\bfq_2-\bfq_1$. The movement of the centre of mass is
irrelevant, and we can concentrate on the relative coordinate
$\bfq$. We can also assume for simplicity that the final positions are
the same as the initial ones. Then the two paths above correspond to
the paths in relative position space depicted in Figure 7.
\begin{figure}[hb]
\epsfysize=5cm
\centerline{\epsfbox{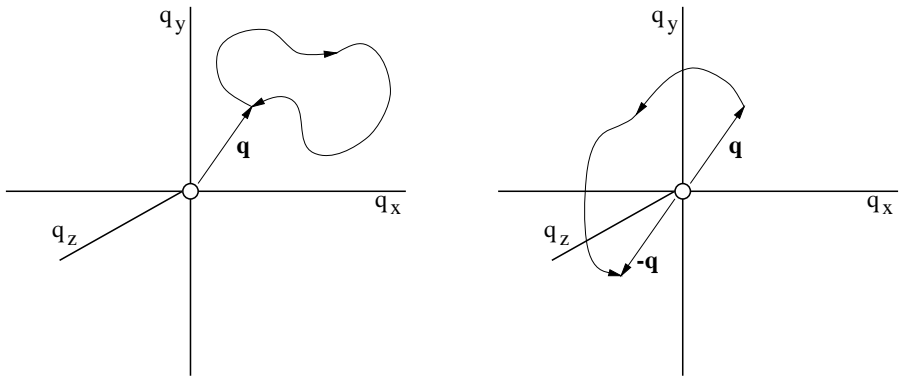}}
\caption{Paths in relative coordinate space.}
\end{figure}

The point is, of course, that the relative positions $\bfq$ and
$-\bfq$ represent the same configuration: interchanging $\bfq_1$ and
$\bfq_2$ changes $\bfq\to-\bfq$.

We can elevate somewhat the tone of the discussion by introducing some
amount of formalism. The configuration space for the relative position
of two identical particles is {\em not} $\bfR^3-\{0\}$,\footnote{Recall
  that we have supposed that the particles have an infinite,
  short-range repulsion; hence the subtraction of the origin (which
  represents coincident points).} as one would have naively thought,
but $(\bfR^3-\{0\})/\bfZ_2$. The
division by the factor $\bfZ_2$ indicates that opposite points in the
space $\bfR^3-\{0\}$, namely any point $\bfq$ and the point
diametrically opposite to it $-\bfq$, are to be {\em
identified}: they represent the same configuration. We must keep this
in mind when we attempt to draw paths: the second path of Figure 7
is a {\em closed} one.

A topological space such as our configuration
space can be characterized as {\em simply connected} or
as {\em non-simply connected} according to whether all paths starting
and finishing at the same point can or cannot be contracted into the
trivial path (representing no relative motion of the particles). It is
clear from Figure 7 that the first path can be deformed to the
trivial path, while the second one cannot, so the configuration space
is not simply connected. Clearly any path which does not correspond to
an exchange of the particles (a ``direct'' path) is topologically
trivial, while any ``exchange'' path is not; we can divide the space
of paths on the configuration space into two topological classes (direct
and exchange). 

Our configuration space is more precisely described as
{\em doubly-connected}, since any two topologically nontrivial (exchange)
paths taken one after the other result in a direct path, which is
trivial. Thus the classes of paths form the elements of the
group $\bfZ_2$ if we
define the product of two paths to mean first one path followed by the
other (a definition which extends readily to the product of classes).

One final bit of mathematical nomenclature: our configuration space,
as noted above, is $(\bfR^3-\{0\})/\bfZ_2$, which is not simply
connected. We define the {\em (simply connected) covering space} as
the simply connected space which looks locally like the original
space. In our case, the covering space is just $\bfR^3-\{0\}$.

At this point you might well be wondering:
what does this have to do with PIs? We can rewrite the PI expression
for the amplitude as the following PI in the covering space
$\bfR^3-\{0\}$:
\bea 
A(\bfq,T;\bfq,0)&=&\sum_{\mathrm{direct}}e^{iS[\bfq]}+
\sum_{\mathrm{exchange}}e^{iS[\bfq]}\nonumber\\
&=&\bar A(\bfq,T;\bfq,0)+\bar A(-\bfq,T;\bfq,0).
\label{stat1}
\eea
The notation $\bar A$ is used to indicate that these PIs are
in the covering space, while $A$ is a PI in the
configuration space. The first term is the sum over all paths from
$\bfq$ to $\bfq$; the second is that for paths from $\bfq$ to $-\bfq$.

Notice that each sub-path integral is a perfectly respectable PI in
its own right: each would be a complete PI for the same
dynamical problem but involving
distinguishable particles. Since the PI can
be thought of as a technique for obtaining the propagator in quantum
mechanics, and since (as was mentioned at the end of Section 2) the
propagator is a solution of the Schroedinger equation, either of these 
sub-path integrals also satisfies it. It follows that we can
generalize the amplitude $A$ to the following expression, which {\em
  still} satisfies the Schroedinger equation:
\bea
A(\bfq,T;\bfq,0)\to A^\phi(\bfq,T;\bfq,0)
&=&\sum_{\mathrm{direct}}e^{iS[\bfq]}+
e^{i\phi}\sum_{\mathrm{exchange}}e^{iS[\bfq]}\nonumber\\
&=&\bar A(\bfq,T;\bfq,0)+e^{i\phi}\bar A(-\bfq,T;\bfq,0).
\label{stat2}
\eea
This generalization might appear to be {\it ad hoc} and ill-motivated,
but we will see shortly that it is intimately related to particle
statistics.

There is a restriction on the added
phase, $\phi$. To see this, suppose that
we no longer insist that the path be a closed one from $\bfq$ to
$\bfq$. Then \eqref{stat2} generalizes to
\beq A^\phi(\bfq',T;\bfq,0)=\bar A(\bfq',T;\bfq,0)
+e^{i\phi}\bar A(-\bfq',T;\bfq,0).
\label{stat2a}
\eeq
If we vary $\bfq'$ continuously to the point $-\bfq'$, we have
\beq
A^\phi(-\bfq',T;\bfq,0)=\bar A(-\bfq',T;\bfq,0)
+e^{i\phi}\bar A(\bfq',T;\bfq,0).
\label{stat2b}
\eeq
But since the particles are identical,
the new final configuration $-\bfq'$ is identical to old one $\bfq'$.
\eqref{stat2a} and \eqref{stat2b} are expressions for the amplitude for
the same physical process, and can differ at most by a phase:
\[ A^\phi(\bfq',T;\bfq,0)=e^{i\alpha}A^\phi(-\bfq',T;\bfq,0).
\]
Combining these three equations, we see that
\[
\bar A(\bfq',T;\bfq,0)+e^{i\phi}\bar A(-\bfq',T;\bfq,0)
=e^{i\alpha}\left(
\bar A(-\bfq',T;\bfq,0)+e^{i\phi}\bar A(\bfq',T;\bfq,0)\right).
\]
Equating coefficients of the two terms, we have $\al=\phi$ (up to a
$2\pi$ ambiguity), and
\[
e^{i2\phi}=1.
\]
This equation has two physically distinct solutions: $\phi=0$ and
$\phi=\pi$. (Adding $2n\pi$ results in physically equivalent
solutions.)

If $\phi=0$, we obtain
\beq
A(\bfq,T;\bfq,0)=\bar A(\bfq,T;\bfq,0)+\bar A(-\bfq,T;\bfq,0),
\label{stat3}
\eeq
the naive sum of the direct and exchange
amplitudes, as is appropriate for Bose statistics.

If, on the other hand, $\phi=\pi$, we obtain
\beq
A(\bfq,T;\bfq,0)=\bar A(\bfq,T;\bfq,0)-\bar A(-\bfq,T;\bfq,0).
\label{stat4}
\eeq
The direct and exchange amplitudes contribute with a relative minus
sign. This case describes Fermi statistics. 

In three dimensions, we see that the PI gives us an elegant way of
seeing how these two types of quantum statistics arise.

\subsubsection{Two dimensions}

We will now repeat the above analysis in two dimensions, and
will see that the difference is significant.

Consider a system of two identical particles in two dimensions, again
adding a short-range, infinitely strong repulsion.
Once again, we restrict
ourselves to the centre of mass frame, since centre-of-mass
motion is irrelevant to the present discussion.
The amplitude that two particles starting at relative position
$\bfq=(q_x,q_y)$ will propagate to a final relative position
$\bfq'=(q_x',q_y')$ in time $T$ is
\[ A(\bfq',T;\bfq,0)=\sum_{\mathrm{paths}}e^{iS[\bfq_1(t),\bfq_2(t)]},
\]
the sum being over all paths from $\bfq$ to $\bfq'$ in the
configuration space. 

Once again, the PI separates into distinct topological classes, but
there are now an infinity of possible classes. To see this, consider
the three paths depicted in Figure 8,
where for simplicity we restrict to the
case where the initial and final configurations are the same. It
is important to remember that drawing paths in the plane is somewhat
misleading: as in three dimensions, 
opposite points are identified, so that a path
from any point to the diametrically opposite point is closed.

\begin{figure}[hb]
\epsfysize=5cm
\centerline{\epsfbox{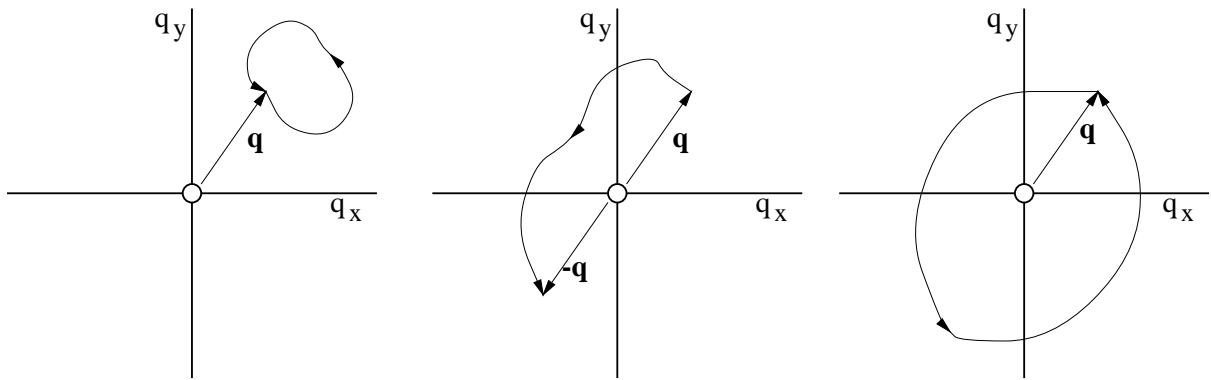}}
\caption{Three topologically distinct paths in two dimensions.}
\end{figure}

The first and second paths are similar to the direct and exchange
paths of the three-dimensional problem. The third path, however,
represents a distinct class of path in two dimensions. The particles
circle around each other, returning to their starting points.

It is perhaps easier to visualize these paths in a three-dimensional
space-time plot, where the vertical axis represents time and the
horizontal axes represent space (Figure 9).

\begin{figure}[hb]
\epsfysize=5cm
\centerline{\epsfbox{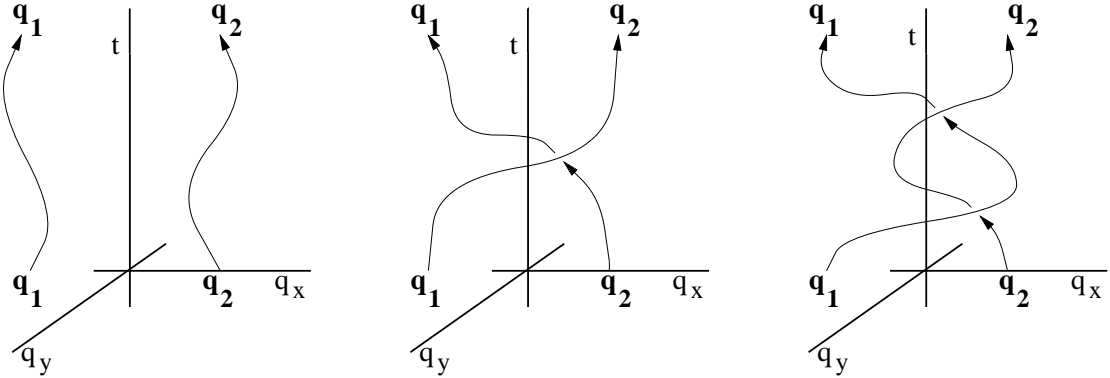}}
\caption{Spacetime depiction of the 3 paths in Figure 8.}
\end{figure}

It is clear that in the third path the particle initially at $\bfq_1$ 
returns to $\bfq_1$, and similarly for the other particle: this path
does not involve a permutation of the particles. It is also
clear that this path cannot be continuously deformed
into the first path, so it is in a distinct topological class. (It
is critical here that we have excised the origin in relative
coordinates -- \ie that we have disallowed configurations where the two
particles are at the same point in space.)

The existence of this third class of paths generalizes in an obvious
way, and we are led to the following conclusion: the paths starting and
finishing at relative
position $\bfq$ can be divided into an infinite
number of classes of paths in the
plane (minus the origin); a class is specified by the number
of interchanges of
the particles (keeping track of the sense of each interchange). This
is profoundly different from the three-dimensional case, where there
were only two classes of paths: direct and exchange.

If we characterize
a path by the polar angle of the relative coordinate,
this angle is $n\pi$ in the $n^{th}$ class, where $n$ is an
integer. (For the three paths shown above, $n=$ 0, 1, and 2,
respectively.)

We can write
\[
A(\bfq,T;\bfq,0)=\sum_{n=-\infty}^{\infty}{\bar A}_n(\bfq,T;\bfq,0),
\]
where ${\bar A}_n$ is the covering-space
PI considering only paths of change of polar
angle $n\pi$.

This path integral can again be generalized to
\[
A(\bfq,T;\bfq,0)=
\sum_{n=-\infty}^{\infty}C_n{\bar A}_n(\bfq,T;\bfq,0),
\]
$C_n$ being phases. Since each ${\bar A}_n$ satisfies the Schroedinger
equation, so does this generalization. 

Again, a restriction on the
phases arises, as can be seen by the following argument. Let us relax
the condition that the initial and final points are the same; let us
denote the
final point $\bfq'$ by its polar coordinates
$(q',\theta')$.
Writing $A(q',\theta')\equiv A(\bfq',T;\bfq,0)$, we have
\[
A(q',\theta')=\sum_{-\infty}^{\infty}C_n{\bar A}_n(q',\theta').
\]
Now, we can change continuously $\theta'\to\theta'+\pi$, yielding
\beq
A(q',\theta'+\pi)=\sum_{n=-\infty}^{\infty}C_n{\bar A}_n(q',\theta'+\pi).
\label{stat4a}
\eeq
Two critical observations can now be made.
First, the final configuration is unchanged, so 
$A(q',\theta'+\pi)$ can differ from $A(q',\theta')$ by at most a phase:
\[
A(q',\theta'+\pi)=e^{-i\phi}A(q',\theta').
\]
Second, ${\bar A}_n(q',\theta'+\pi)=
{\bar A}_{n+1}(q',\theta')$, since this is just two different ways of
expressing exactly the same quantity. Applying these two observations to
\eqref{stat4a},
\bea
e^{-i\phi}\sum_{n=-\infty}^{\infty}C_n{\bar A}_n(q',\theta')
&=&\sum_{-\infty}^{\infty}C_n{\bar A}_{n+1}(q',\theta')\nonumber\\
&=&\sum_{-\infty}^{\infty}C_{n-1}{\bar A}_n(q',\theta').
\eea
Equating coefficients of ${\bar A}_n(q',\theta')$, we get
\[ C_n=e^{i\phi}C_{n-1}.
\]
Choosing $C_0=1$, we obtain for the amplitude
\beq
A=\sum_{-\infty}^{\infty}e^{in\phi}\bar A_n,
\label{stat6}
\eeq
which is the two-dimensional analog of the three-dimensional result
\eqref{stat2}.

The most important observation to be made is that
there is no longer a restriction on the angle $\phi$, as was the case
in three dimensions. We see that, relative to the ``naive'' PI (that
with $\phi=0$), the class corresponding to a net number $n$ of
counter-clockwise rotations of one particle around the other
contributes with an extra phase $\exp in\phi$. If $\phi=0$ or $\pi$,
this collapses to the usual cases of Bose and Fermi statistics,
respectively. However in the general case the phase relation between
different paths is more complicated (not determined by whether the
path is ``direct'' or ``exchange''); this new possibility is known as
fractional statistics, and particles obeying these statistics are
known as anyons.

Anyons figure prominently in the accepted theory of the fractional
quantum Hall effect, and were proposed as being relevant to
high-temperature superconductivity, although that possibility seems not
to be borne out by experimental results. Perhaps Nature has other
applications of fractional statistics which await discovery.

\subsection{Magnetic Monopoles and Charge Quantization}

All experimental evidence so far tells us
that all particles have electric charges which
are integer multiples of a fundamental unit of electric charge,
$e$.\footnote{The unit of electric charge is more properly $e/3$,
that of the quarks; for simplicity, I will ignore this fact.}
There is absolutely nothing wrong with a theory of electrodynamics of
particles of arbitrary charges: we could have particles of charge $e$
and $\sqrt{17}e$, for example. It was a great mystery why charge was
quantized in the early days of quantum mechanics.

In 1931, Dirac showed that the quantum mechanics of charged particles
in the presence of magnetic monopoles is problematic, unless the
product of the electric and magnetic charges is an integer multiple of
a given fundamental value. Thus, the existence of monopoles implies
quantization of electric charge, a fact which has fueled experimental
searches for and theoretical speculations about magnetic monopoles
ever since.

We will now recast Dirac's argument into a modern form in terms of PIs.
A monopole of charge $g$ positioned at the origin has magnetic field
\[  \bfB=g{{\hat\bfe}_r\over r^2}.
\]
As is well known,
this field cannot be described by a normal (smooth, single-valued)
vector potential: writing $\bfB=\nabla\times\bfA$ implies that the
magnetic flux emerging from any closed surface (and thus the
magnetic charge contained in any such surface) must be zero. This fact
makes life difficult for monopole physics, for several
reasons. Although classically the Maxwell equations and the Lorentz
force equation form a complete set of equations for particles (both
electrically and magnetically charged, with the simple addition of
magnetic source terms) and
electromagnetic fields, their derivation from an action principle
requires that the electromagnetic field be described in terms of the
electromagnetic potential, $A_\mu$. Quantum mechanically, things are
even more severe: one cannot avoid $A_\mu$,
because the coupling of a particle to the electromagnetic
field is written in terms of $A_\mu$, not the electric and magnetic
fields.

Dirac suggested that if a monopole exists, it could be described by an
infinitely-thin and tightly-wound solenoid carrying a magnetic flux
equal to that of the monopole. The solenoid is semi-infinite in
length, running from the position of the monopole to infinity along an
arbitrary path. The magnetic
field produced by such a solenoid can be shown to be that of a
monopole plus the usual field produced by a solenoid, in this case,
an infinitely intense, infinitely narrow tube of flux running from the
monopole to infinity along the position of the solenoid
(Figure 10). Thus except inside the solenoid,
the field produced is that of the monopole. The field inside the
solenoid is known as the ``Dirac string''.
\begin{figure}[hb]
\epsfysize=7cm
\centerline{\epsfbox{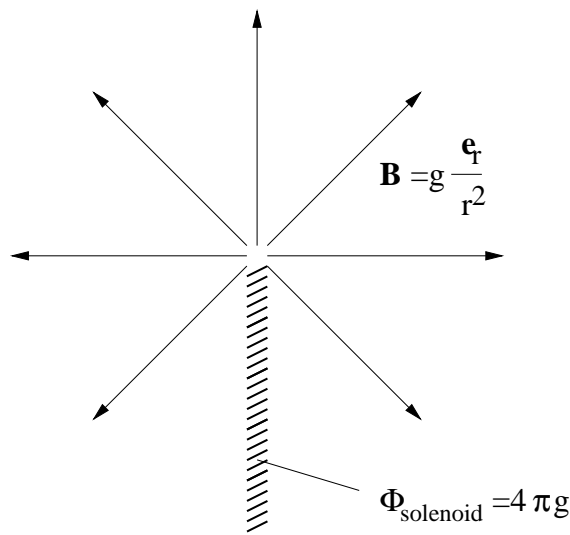}}
\caption{Monopole as represented by a semi-infinite, infinitely
  tightly-wound solenoid.}
\end{figure}
The flux brought into any closed surface including the monopole is
now zero, because the solenoid brings in a flux equal to that
flowing out due to the monopole. Thus, the combined monopole-solenoid
can be described by a vector potential.

However, in order for this to be a valid description of the monopole,
we must somehow convince
ourselves that the solenoid can be made invisible to any electrically
charged particle
passing by it. We can describe the motion of
such a particle by a PI, and
two paths passing on either side of the Dirac string can contribute to
the PI (Figure 11). But the vector potential of the Dirac string will
affect the action of each of these paths differently, as we have seen
in the Aharonov-Bohm effect.
\begin{figure}[hb]
\epsfysize=4cm
\centerline{\epsfbox{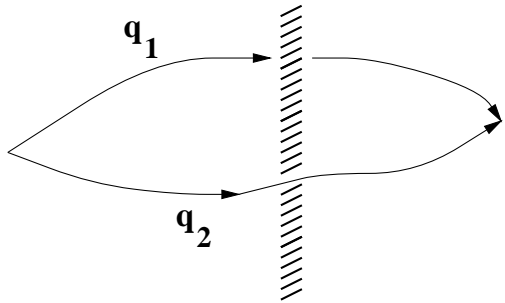}}
\caption{Paths contributing to the propagator in the presence of a
  monopole. The paths form a loop encircling the Dirac string.}
\end{figure}

In order that the interference of these paths be unaffected by the
presence of the Dirac string, the relative phase must be an integral
multiple of $2\pi$. This phase is
\[ -{e\over\hbar c}\oint d\bfq\cdot\bfA=-{e\Phi\over\hbar c}
=-{4\pi eg\over\hbar c}.
\]
Setting this to $2\pi n$, in order for the motion of a particle of
charge $e$ to be unaffected by the presence of the Dirac string, the
electric charge must be
\beq
e={2\pi\hbar c\over 4\pi g}n={\hbar c\over2g}n.
\label{quantcond}
\eeq
Thus, the existence of magnetic monopoles requires the quantization of
electric charge; the fundamental unit of electric charge is $2\pi\hbar
c/g$.

In modern theories of fundamental physics, Grand Unified Theories also
imply quantization of electric charge, apparently avoiding the
necessity for magnetic monopoles. But any Grand Unified Theory
actually has magnetic
monopoles as well (though they are of a nature quite different to the
``Dirac monopole''), so the intimate relation between magnetic
monopoles and the quantization of electric charge is preserved, albeit
in a form quite different from that suggested by Dirac.

\newpage\thispagestyle{empty}
\section[Statistical Mechanics]
{Statistical Mechanics via Path Integrals}

The path integral turns out to provide an elegant way of
doing statistical mechanics. The reason for this is that, as we will
see, the central object in statistical mechanics, the partition
function, can be written as a PI. Many books have been written
on statistical mechanics with emphasis on path integrals,
and the objective in this lecture is simply to see
the relation between the partition function and the PI.

The definition of the partition function is
\beq
Z=\sum_j e^{-\be E_j},
\label{sm1}
\eeq
where $\be=1/\kbt$ and
$E_j$ is the energy of the state $\ket j$. We can write
\[ Z=\sum_j\bra j e^{-\be H}\ket j = \Tr e^{-\be H}.
\]
But recall the definition of the propagator:
\[ K(q',T;q,0)=\bra{q'}e^{-iHT}\ket{q}.
\]
Suppose we consider $T$ to be a complex parameter, and consider it to
be pure imaginary, so that we can write $T=-i\be$, where $\be$ is
real. Then
\bea
K(q',-i\be;q,0)&=&\bra{q'}e^{-iH(-i\be)}\ket{q}\nonumber\\
&=&\bra{q'}e^{-\be H}\underbrace{\sum_j\ket{j}\bra{j}}_{=1}
\ket{q}\nonumber\\
&=&\sum_j e^{-\be E_j}\bra{q'}j\rangle\langle j\ket{q}
\nonumber\\
&=&\sum_j e^{-\be E_j}\langle j\ket{q}\bra{q'}j\rangle.
\nonumber
\eea
Putting $q'=q$ and integrating over $q$, we get
\beq
\int dq \,K(q,-i\be;q,0)=
\sum_j e^{-\be E_j}\bra{j}
\underbrace{\int dq\ket{q}\bra{q}}_{=1}\ket{j}=Z.
\label{sm2}
\eeq
This is the central observation of this section: that the
propagator evaluated at negative imaginary time is related to the
partition function.

We can easily work out an elementary example such as the harmonic
oscillator. Recall the path integral for it, \eqref{harmosc2}:
\[
K(q',T;q,0)=\left({m\om\over2\pi i\sin\om T}\right)^{1/2}
\exp \left\{i {m\om\over2\sin\om T}
\left(({q'}^2+q^2)\cos\om T-2q'q\right)\right\}.
\]
We can put $q'=q$ and $T=-i\be$:
\[ K(q,-i\be;q,0)=\left({m\om\over2\pi \sinh(\be\om)}\right)^{1/2}
\exp \left\{- {m\om q^2\over\sinh(\be\om)}
\left(\cosh(\be\om)-1\right)\right\}.
\]
The partition function is thus
\bea
Z&=&\int dq\,K(q,-i\be;q,0)
=\left({m\om\over2\pi \sinh(\be\om)}\right)^{1/2}
\sqrt{\pi\over{m\om \over\sinh(\be\om)}
\left(\cosh(\be\om)-1\right)}
\nonumber\\
&=&\left[2(\cosh(\be\om)-1)\right]^{-1/2}
=\left[e^{\be\om/2}(1-e^{-\be\om})\right]^{-1}
\nonumber\\
&=&{e^{-\be\om/2}\over 1-e^{-\be\om}}
=\sum_{j=0}^\infty e^{-\be(j+1/2)\om}.\nonumber
\eea
Putting $\hbar$ back in, we get the familiar result
\[
Z=\sum_{j=0}^\infty e^{-\be(j+1/2)\hbar\om}.
\]

The previous calculation actually had nothing to do with PIs. The
result for $K$ was derived \via PIs earlier, but it can be derived
(more easily, in fact) in ordinary quantum mechaincs.
However we can
rewrite the partition function in terms of a PI. In ordinary (real)
time,
\[
K(q',T;q,0)=\int\DD q(t)
\exp i\int_0^T dt\left({m{\dot q}^2\over2}-V(q)\right),
\]
where the integral is over all paths from $(q,0)$ to $(q',T)$.
With $q'=q$, $T\to-i\be$,
\[
K(q,-i\be;q,0)=\int\DD q(t)
\exp i\int_0^{-i\be} dt\left({m{\dot q}^2\over2}-V(q)\right).
\]
where we now integrate along the negative imaginary time axis (Figure 12).

\begin{figure}[hb]
\epsfysize=5cm
\centerline{\epsfbox{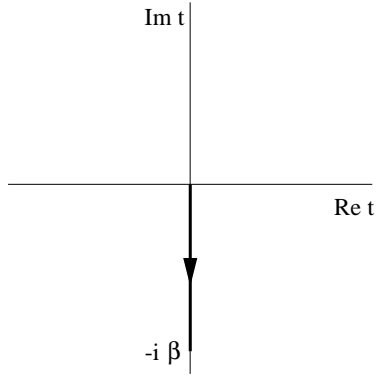}}
\caption{Path in the complex time plane.}
\end{figure}

Let us define a real variable for this integration, $\tau=it$. $\tau$
is called the imaginary time, since when the time $t$ is imaginary,
$\tau$ is real. (Kind of confusing, admittedly, but true.)
Then the integral over $\tau$ is along {\em its} real axis: when
$t:0\to-i\be$, then $\tau:0\to\be$. We can write $q$ as a function of
the variable $\tau$: $q(t)\to q(\tau)$; then $\dot q = i dq/d\tau$.
The propagator becomes
\beq
K(q,-i\be;q,0)=\int\DD q(\tau)
\exp -\int_0^\be d\tau\left({m\over2}\left({dq\over d\tau}\right)^2
+V(q)\right).
\label{sm3}
\eeq
The integral is over all functions $q(\tau)$ such that
$q(0)=q(\be)=q$.

The result \eqref{sm3} is an ``imaginary-time'' or ``Euclidean'' path
integral, defined by associating to each path an amplitude
(statistical weight) $\exp - S_E$, where $S_E$ is the so-called
Euclidean action, obtained from the usual (``Minkowski'') action
by changing the sign of the potential energy term.

The Euclidean PI might seem like a strange, unphysical
beast, but it actually has many uses. One will be discussed in the
next section, where use will be made of the fact that at low
temperatures the ground state gives the dominant contribution to the
partition function. It can therefore be used to find the ground state
energy. We will also see the Euclidean PI in Section 9, when
discussing the subject of instantons, which are used to describe
phenomena such as quantum mechanical tunneling.

\newpage\thispagestyle{empty}
\section[Perturbation Theory I]
{Perturbation Theory in Quantum Mechanics}

We can use the Euclidean PI to compute a perturbation expansion for
the ground state energy (among other things). This is not terribly
useful in and of itself (once again, conventional techniques are
a good deal easier), but the techniques used
are very similar to those used in perturbation theory
and Feynman diagrams in field theory. For this reason, we will discuss
corrections to the ground state energy of an elementary quantum
mechanical system in some detail.

From $Z$ it is quite easy to extract the ground state energy. (This is
a well-known fact of statistical mechanics, quite independent of PIs.) 
From the definition of
$Z$,
\[ Z(\be)=\sum_j e^{-\be E_j},
\]
we can see that the contribution of each state decreases exponentially
with $\be$. However, that
of the ground state decreases less slowly
than any other state. So in the limit of large $\be$ (\ie low
temperature), the ground state contribution will dominate. (This is
mathematically straightforward, and also physically reasonable.) One
finds
\beq
E_0=-\lim_{\be\to\infty}{1\over\be}\log Z.
\label{pt1}
\eeq
In fact, we can extract $E_0$ from something slightly easier to
calculate than $Z$. Rather than integrating over the initial (= final)
position, as with $Z$, let us look at the Euclidean
propagator from $q=0$ to $q'=0$ (the choice of zero is arbitrary).
\[
K_E(0,\be;0,0)=\bra{q'=0}e^{-\be H}\ket{q=0}.
\]
We can insert a complete set of eigenstates of $H$:
\bea
K_E(0,\be;0,0)&=&\bra{q'=0}e^{-\be H}
\sum_j\ket{j}\bra{j}q=0\rangle\nonumber\\
&=&\sum_j e^{-\be E_j}\phi_j(0)\phi_j^*(0),\nonumber
\eea
where $\phi_j$ are the wave functions of $H$. Again the ground state
dominates as $\be\to\infty$, and
\beq 
E_0=-\lim_{\be\to\infty}{1\over\be}\log K_E(0,\be;0,0).
\label{pt1a}
\eeq
(As $\be\to\infty$, the difference between $\be^{-1}\log Z$ and
$\be^{-1}\log K_E$ goes to zero.)

So let us see how we can calculate $K_E(0,\be;0,0)$ perturbatively via
the PI. The starting point is
\[
K_E(0,\be;0,0)=\int\DD q\,e^{-S_E(\dot q,q)},
\]
where the paths over which we integrate start and finish at $q=0$, and
where the Euclidean action is
\[
S_E=\int_0^\be d\tau\left({m{\dot q}^2\over2}+V(q)\right),
\]
and with $\dot q=dq/d\tau$. As an example, consider the anharmonic
oscillator, with quadratic and quartic terms in the potential:
\beq
K_E(0,\be;0,0)=\int\DD q\,\exp {-\int d\tau
\left(\half m{\dot q}^2+\half m\om^2q^2+{\la\over4!} q^4\right)}.
\label{pt2}
\eeq

Clearly it is the quartic term which complicates life considerably; we
cannot do the PI exactly.\footnote{In fact, 
the situation is exactly like the
evaluation of the ordinary integral
\[ I=\int_{-\infty}^\infty dx\,\exp{-(\half x^2+{\la\over4!}x^4)},
\]
which looks innocent enough but which cannot be evaluated exactly.
The technique which we will
develop to evaluate \eqref{pt2} can also be used for this
ordinary integral -- an amusing and recommended exercise.}
But we can use the following trick to evaluate it perturbatively in
$\la$. (This trick is far more complicated than necessary for this
problem, but is a standard -- and necessary! -- trick in quantum field
theory.) Define $K_E^0[J]$, the PI for a harmonic oscillator with a
source term (which describes the action of an external force) added 
to the Lagrangian:
\beq
K_E^0[J]=\int\DD q\,\exp {-\int d\tau
\left(\half m{\dot q}^2+\half m\om^2q^2-J(\tau)q(\tau)\right)}.
\label{pt3}
\eeq
Unlike \eqref{pt2},
this PI can be evaluated exactly; we will do this (as much as is
necessary, at least) shortly. Once we have evaluated it, how does it
help us to compute \eqref{pt2}? To see the use of $K_E^0[J]$, acting
on it with a derivative has the effect of putting a factor $q$ in the
PI: for any time $\tau_1$,
\[
{\delta K_E^0[J]\over\delta J(\tau_1)}
=\int\DD q\,q(\tau_1)\exp {-\int d\tau
\left(\half m{\dot q}^2+\half m\om^2q^2-J(\tau)q(\tau)\right)}.
\]
A second derivative puts a second $q$ in the PI:
\[
{\delta^2 K_E^0[J]\over\delta J(\tau_1)\delta J(\tau_2)}
=\int\DD q\,q(\tau_1)q(\tau_2)\exp {-\int d\tau
\left(\half m{\dot q}^2+\half m\om^2q^2-J(\tau)q(\tau)\right)}.
\]
In fact, we can generalize this to an arbitrary functional $F$:
\beq
F\left[\delta\over\delta J\right] K_E^0[J]
=\int\DD q\,F[q]e^{- S_E^0[J]},
\label{pt4}
\eeq
where $S_E^0[J]$ is the Euclidean action for the harmonic oscillator
with source. 
(To prove \eqref{pt4}, bring $F[\delta/\delta J]$ 
inside the PI; each
$\delta/\delta J$ in $F$ operating on $\exp- S_E^0[J]$ gives rise
to a $q$ in front of the exponential.)

Now, if we choose $F[q]=\exp-\int d\tau{\la\over4!}q^4$, we get:
\bea
e^{-\int d\tau{\la\over4!}
\left(\delta\over\delta J\right)^4}
K_E^0[J]
&=&\int\DD q\,\exp \left\{{-\int d\tau{\la\over4!} q^4}\right\}
e^{- S_E^0[J]}\nonumber\\
&=&\int\DD q\,\exp{-\int d\tau
\left(\half m{\dot q}^2+\half m\om^2q^2
+{\la\over4!} q^4-J(\tau)q(\tau)\right)}.\nonumber
\eea
If we {\em now} put $J=0$, we have the PI we started with. So the
final result is:
\beq
\fbox{$\displaystyle
K_E(0,\be;0,0)=\left.\left(
\exp\left\{{-\int d\tau{\la\over4!}
\left(\delta\over\delta J\right)^4}\right\}
K_E^0[J]
\right)\right|_{J=0}.
$}
\label{pt5}
\eeq

We can, and will, calculate $K_E^0[J]$ as an explicit functional of
$J$. If we then expand the exponential which operates on it in 
\eqref{pt5}, we get a power series in $\la$:
\bea
K_E(0,\be;0,0)&=&\left\{\left(1-\int d\tau{\la\over4!}
\left(\delta\over\delta J(\tau)\right)^4\right.\right.\nonumber\\
&&\qquad\left.\left.\left.+{1\over2!}\int d\tau{\la\over4!}
\left(\delta\over\delta J(\tau)\right)^4\int d\tau'{\la\over4!}
\left(\delta\over\delta J(\tau')\right)^4+\cdots\right)
K_E^0[J]\right\}\right|_{J=0}\nonumber\\
&=&K_E^0[J]-{\la\over4!}\left.\left(\int d\tau
\left(\delta\over\delta J(\tau)\right)^4 K_E^0[J]\right)
\right|_{J=0}+o(\la^2).\nonumber
\eea

Let us now evaluate $K_E^0[J]$,
\[
K_E^0[J]=\int\DD q\,\exp {-\int d\tau
\left(\half m{\dot q}^2+\half m\om^2q^2-J(\tau)q(\tau)\right)}.
\]
To do this, suppose that we can find the {\em classical path}
$q_{cJ}(\tau)$, the solution of
\beq
m\ddot q=m\om^2q-J(\tau),\qquad q(0)=q(\beta)=0.
\label{pt6}
\eeq
Once we have done this, we can perform a change of variables in the PI:
we define $q(\tau)=q_{cJ}(\tau)+y(\tau)$, and integrate over paths
$y(\tau)$. This is useful because
\bea
\int d\tau\left(\half m{\dot q}^2
+\half m\om^2 q^2 -J(\tau)q(\tau)\right)
&=&\int d\tau\left(\half m{\dot q}_{cJ}^2
+\half m\om^2q_{cJ}^2-J(\tau){q_{cJ}}(\tau)\right)\nonumber\\
&&+\underbrace{(\mbox{linear in $y$})}_{=0}
+\int d\tau\left(\half m {\dot y}^2
+\half m\om^2y^2\right).\nonumber
\eea
The linear term vanishes because $q_{cJ}$ satisfies
the equation of motion. So the PI becomes
\[
K_E^0[J]=e^{-S_{Ec}[J]}\int\DD y\,
\exp {-\int d\tau\left(\half m{\dot y}^2+\half m\om^2y^2\right)}.
\]
The crucial observation is that the resulting PI is {\em independent
  of $J$}: it is an irrelevant constant; call it $C$. (In fact, $C$
  is neither constant [it depends on $\be$], nor entirely irrelevant
  [it is related to the unperturbed ground state energy, as we will
  see]. Crucial for the present purposes is that $C$ is independent of
  $J$.)
\[
K_E^0[J]=C\,e^{-S_{Ec}[J]},
\]
where
\[ S_{Ec}[J]=\int d\tau\left(\half m{\dot q}_{cJ}^2
+\half m\om^2q_{cJ}^2-J(\tau){q_{cJ}}(\tau)\right).
\]
This can be simplified by integrating the first term by parts,
yielding
\[ S_{Ec}[J]=\int d\tau\,q_{cJ}\left(-\half m{\ddot q}_{cJ}
+\half m\om^2{q_{cJ}}-J(\tau)\right).
\]
Using the classical equation of motion \eqref{pt6}, we get
\[
S_{Ec}[J]=-\half\int d\tau\,J(\tau)q_{cJ}(\tau).
\]

We must still solve the classical problem, \eqref{pt6}. The solution
can be written in terms of the Green's function for the problem. Let
$G(\tau,\tau')$ be the solution of
\bea
m\left({d^2\over d\tau^2}-\om^2\right)G(\tau,\tau')
&=&\delta(\tau-\tau'),\nonumber\\
G(0,\tau')=G(\be,\tau')&=&0.\nonumber
\eea
Then we can immediately write
\[
q_{cJ}(\tau)=\int_0^\be d\tau' G(\tau,\tau')J(\tau'),
\]
which can be proven by substution into \eqref{pt6}. We can now write
\beq
\fbox{$\displaystyle
K_E^0[J]=C\,\exp
{\half\int d\tau d\tau'\,J(\tau)G(\tau,\tau')J(\tau')}.
$}
\label{pt8}
\eeq

We can find the Green's function easily in the limit
$\be\to\infty$. It is slightly more convenient to treat the initial
and final times more symmetrically, so let us choose the time interval
to be $(-\be/2,+\be/2)$; in the limit $\be\to\infty$ we go from
$-\infty$ to $\infty$. Then we have
\[ m\left({d^2\over d\tau^2}-\om^2\right)G(\tau,\tau')
=\delta(\tau-\tau').
\]
By taking the Fourier transform, we see that
\beq
G(\tau,\tau')=-{1\over m}\int_{-\infty}^\infty
{dk\over2\pi}{1\over(k^2+\om^2)}
e^{ik(\tau-\tau')}.
\label{gf}
\eeq

We can now compute the first-order correction to $K_E$ (from which we
get the first-order correction to the ground state energy). We have
\beq
K_E=K_E^0[0]-{\la\over4!}\left.\int d\tau
\left({\delta\over\delta J(\tau)}\right)^4 K_E^0[J]\right|_{J=0}.
\label{pt8a}
\eeq
Since in the second term
we take four derivatives of $K_E^0[J]$ and then set $J=0$, only
the piece of  $K_E^0[J]$ which is quartic in $J$ is relevant: fewer
than four $J$'s will be killed by the derivatives; more than four will
be killed when setting $J=0$.
\bea
K_E^0[J]&=&C\,\exp \half\int d\tau d\tau'\,
J(\tau)G(\tau,\tau')J(\tau')\nonumber\\
&=&\mathrm{irrelevant}+C\cdot\half\left(\half
\int d\tau d\tau'\,J(\tau)G(\tau,\tau')J(\tau')\right)^2
\nonumber\\
&=&{C\over 8}\vev{J_1 G_{12} J_2}\vev{J_3 G_{34} J_4},\label{pt9}
\eea
where we have used the compact notation
$\vev{J_1 G_{12} J_2}=
\int d\tau_1 d\tau_2\,J(\tau_1)G(\tau_1,\tau_2)J(\tau_2)$.

Substituting \eqref{pt9} into \eqref{pt8a},
\beq
K_E=C\left(1-{\la\over4!}{1\over8}\int d\tau
\left({\delta\over\delta J(\tau)}\right)^4
\vev{J_1 G_{12} J_2}\vev{J_3 G_{34} J_4}
+o(\la^2)\right).
\label{pt10}
\eeq

To ensure that we
understand the notation and how functional differentiation works, let
us work out a slightly simpler example than the above.
Consider
\[
X\equiv
\left({\delta\over\delta J(\tau)}\right)^2
\vev{J_1 G_{12} J_2}
=\left({\delta\over\delta J(\tau)}\right)^2
\int d\tau_1 d\tau_2\,J(\tau_1)G(\tau_1,\tau_2)J(\tau_2).
\]
The first derivative can act either on $J_1$ or $J_2$. In either case,
it gives a delta function, which will make one of the integrals collapse:
\bea
X&=&{\delta\over\delta J(\tau)}\int d\tau_1 d\tau_2\left(
\delta(\tau-\tau_1)G(\tau_1,\tau_2)J(\tau_2)
+J(\tau_1)G(\tau_1,\tau)\delta(\tau-\tau_2)\right)\nonumber\\
&=&{\delta\over\delta J(\tau)}
\left(\int d\tau_2 G(\tau,\tau_2)J(\tau_2)
+\int d\tau_1 J(\tau_1)G(\tau_1,\tau)\right)\nonumber
\eea
In each term the remaining derivative acts similarly and
kills the remaining integral; the result is
\[ X=2 G(\tau,\tau).
\]

The functional derivatives in \eqref{pt10} are a straightforward
generalization of this; we find
\[
K_E=C\left(1-{1\over8}{\la\over4!}\int d\tau\,4!G(\tau,\tau)^2
\right).
\]
From \eqref{gf} $G(\tau,\tau)=-1/2m\om$; the $\tau$ integral is just
the time interval $\be$, and finally we get
\beq
K_E(0,\be;0,0)
=C\left(1-{\be\la\over32m^2\om^2}+o(\la^2)\right)
=C e^{-\be\la/32m^2\om^2}
\label{pt11}
\eeq
to order $\la$.

Now we can put this expression to good use, extracting the ground
state energy from \eqref{pt1a}:
\[ E_0=-\lim_{\be\to\infty}{1\over\be}\log K_E(0,\be;0,0)
=-\lim_{\be\to\infty}{1\over\be}
\left(\log C-{\be\la\over32m^2\om^2}\right).
\]
Recall that the constant $C$ in \eqref{pt11} depends on $\be$; this
dependence must account for the ground state energy; the term linear
in $\la$ gives the first correction to the energy. Thus
\[ E_0=\half\hbar\om+{\hbar^2\la\over32m^2\om^2}+o(\la^2)
\]
where we have reintroduced $\hbar$. We can check this result against
standard perturbation theory (which is considerably easier!); the
first-order correction to the ground state energy is
\[\int_{-\infty}^\infty
 dq \phi_0^*(q)\left({\la\over4!}q^4\right)\phi_0(q)=\cdots
={\hbar^2\la\over32m^2\om^2},
\]
as above.

One's sanity would be called into question were it suggested that the
PI calculation is a serious competitor for standard perturbation theory,
although the latter itself gets rapidly more and more messy at higher
orders. The technique above also gets messier, but it may well be that
its messiness increases less quickly than that of standard
perturbation theory. If so, the PI calculation could become
competitive with standard perturbation theory at higher orders.
But really the main motivation for discussing the above method is that
it mimics in a more familiar
setting standard perturbation techniques in
quantum field theory.

To summarize this long and somewhat technical
section, let us recall the main features of the
above method. We express the ground state energy as an expression
involving the propagator, \eqref{pt1a}. We separate the Lagrangian into
a ``free'' (\ie quadratic) part and an ``interacting'' (beyond
quadratic) part. Via the PI,
we write the interacting propagator in terms of a
free propagator with source term added, \eqref{pt5}; this expression
is amenable to a perturbation expansion. The free propagator can be
evaluated explicitly, \eqref{pt8}; then \eqref{pt5}
can be computed to any desired order. From this we obtain directly
the ground state energy to the same order.

\newpage\thispagestyle{empty}
\section{Green's Functions in Quantum Mechanics}

In quantum field theory we are interested in objects such as
\[
\bra{0}T\hat\phi(x_1)\hat\phi(x_2)\cdots\hat\phi(x_n)\ket{0},
\]
the vacuum expectation value of a time-ordered product of Heisenberg
field operators. This object is known as a Green's function or as a
correlation function. The order of the operators is such that the
earliest field is written last (right-most), the second earliest
second last, \etc For example,
\[ T\hat\phi(x_1)\hat\phi(x_2)=\left\{\begin{array}{rl}
\hat\phi(x_1)\hat\phi(x_2)&x_1^0>x_2^0\\
\hat\phi(x_2)\hat\phi(x_1)&x_2^0>x_1^0\end{array}\right.
\]
Green's functions are related to amplitudes for physical processes
such as scattering and decay processes. (This point is explained in
most quantum field theory books.)

Let us look at the analogous object in quantum mechanics:
\[ G^{(n)}(t_1,t_2,\cdots,t_n)=\bra{0}
T\hat q(t_1)\hat q(t_2)\cdots\hat q(t_n)\ket{0}.
\]
We will develop a PI expression for this. 

First, we must recast the PI
in terms of Heisenberg representation objects. The operator
$\hat q(t)$ is the usual Heisenberg operator, defined in terms of the
Schroedinger operator $\hat q$ by
$\hat q(t)=e^{iHt}\hat q e^{-iHt}$. The eigenstates of the Heisenberg
operator are $\ket{q,t}$: $\hat q(t)\ket{q,t}=q\ket{q,t}$. The
relation with the time-independent eigenstates is
$\ket{q,t}=e^{iHt}\ket{q}$.\footnote{There is a possible point of
  confusion here. We all know that Heisenberg states are independent of
  time, yet the eigenstates of $\hat q(t)$ depend on time. Perhaps the
  best way to
  view these states $\ket{q,t}$ is that they form, for any fixed time, a
  complete set of states. Just like the usual (time-independent)
  Heisenberg state $\ket{q}$ describes a
  particle which is localized
  at the point $q$ {\em at time $t=0$}, the state $\ket{q,t}$
  describes a particle which is localized
  at the point $q$ {\em at time $t$}.} Then we
can write the PI:
\[
K=\bra{q'}e^{-iHT}\ket{q}=\bra{q',T}q,0\rangle=
\int\DD q\,e^{iS}.
\]

We can now calculate the ``2-point function'' $G(t_1,t_2)$, via the
PI. We will proceed in two steps.
First, we will calculate the following expression:
\[
\bra{q',T}T\hat q(t_1)\hat q(t_2)\ket{q,0}.
\]
We will then devise a method for extracting the vacuum contribution to
the initial and final states.

Suppose first that $t_1>t_2$. Then
\beano
\bra{q',T}T\hat q(t_1)\hat q(t_2)\ket{q,0}
&=&\bra{q',T}\hat q(t_1)\hat q(t_2)\ket{q,0}\\
&=&\int dq_1 dq_2\langle q',T\ket{q_1,t_1}
\underbrace{\bra{q_1,t_1}\hat q(t_1)}_{\bra{q_1,t_1}q_1}
\underbrace{\hat q(t_2)\ket{q_2,t_2}}_{q_2\ket{q_2,t_2}}
\bra{q_2,t_2}q,0\rangle\\
&=&\int dq_1 dq_2\,q_1q_2\langle q',T\ket{q_1,t_1}
\bra{q_1,t_1}q_2,t_2\rangle\bra{q_2,t_2}q,0\rangle.
\eeano
Each of these matrix elements is a PI:
\[ \bra{q',T}T\hat q(t_1)\hat q(t_2)\ket{q,0}=
\int dq_1 dq_2\,q_1q_2
\int_{q_1,t_1}^{q',T}\DD q \,e^{iS}
\int_{q_2,t_2}^{q_1,t_1}\DD q \,e^{iS}
\int_{q,0}^{q_2,t_2}\DD q \,e^{iS}.
\]
This expression consists of a first PI from the initial position $q$
to an arbitrary position $q_2$, a second one from there to a second
arbitrary position $q_1$, and a third one from there to the final
position $q'$. So we are integrating over all paths from $q$ to $q'$,
subject to the restriction that the paths pass through the
intermediate points $q_1$ and $q_2$.
We then integrate over the two arbitrary positions, so that in fact
we are integrating over {\em all} paths:
we can combine these three path integrals plus the integrations over
$q_1$ and $q_2$ into one PI. The factors $q_1$
and $q_2$ in the above
integral can be incorporated into this PI by simply including a factor
$q(t_1)q(t_2)$ in the PI. So
\[
\bra{q',T}\hat q(t_1)\hat q(t_2)\ket{q,0}=
\int_{q,0}^{q',T}\DD q \,q(t_1)q(t_2)e^{iS}\qquad (t_1>t_2).
\]

An identical calculation shows that exactly this same final expression
is also valid for $t_2<t_1$: magically, the PI does the time ordering
automatically. Thus for all times
\[ \bra{q',T}T\hat q(t_1)\hat q(t_2)\ket{q,0}=
\int_{q,0}^{q',T}\DD q \,q(t_1)q(t_2)e^{iS}.
\]

As for how to obtain vacuum-to-vacuum matrix elements, our work on
statistical mechanics provides us with a clue. We can expand the
states $\bra{q',T}$ and $\ket{q,0}$ in terms of
eigenstates of the Hamiltonian. If we evolve towards a
{\em negative imaginary time}, the contribution of all other states will
decay away relative to that of the ground state. We have (resetting
the initial time to $-T$ for convenience)
\[
\bra{q',T}q,-T\rangle\propto\bra{0,T}0,-T\rangle,
\]
where on the right the ``0'' denotes the ground state. The
proportionality involves the ground state wave function and an
exponential factor $\exp2iE_0T=\exp-2E_0|T|$.

We could perform all calculations in a Euclidean theory and
analytically continue to real time when computing physical quantities
(many books do this), but to be closer to physics we can also
consider $T$ not to be pure imaginary and negative, but to have a
small negative imaginary phase: $T=|T|e^{-i\eps}$ ($\eps>0$). {\em In
  what follows, I will simply write $T$, but please keep in mind that
  it has a negative imaginary part!} With this,
\[
\bra{0,T}0,-T\rangle \propto \bra{q',T}q,-T\rangle=
\int\DD q \,e^{iS}.
\]
To compute the Green's functions, we must simply add 
$T\hat q(t_1)\hat q(t_2)\cdots\hat q(t_n)$ to the matrix element, and
the corresponding factor $q(t_1)q(t_2)\cdots q(t_n)$ inside the PI:
\[
\bra{0,T}T\hat q(t_1)\hat q(t_2)\cdots\hat q(t_n)
\ket{0,-T} \propto
\int\DD q \,q(t_1)q(t_2)\cdots q(t_n)e^{iS}.
\]
The proportionality sign is a bit awkward; fortunately,
we can rid ourselves of it. To do this, we note that the left hand
expression is not exactly what we want: the vacua $\ket{0,\pm T}$
differ by a phase. We wish to eliminate this phase; to this end, the
Green's functions are defined
\beano
G^{(n)}(t_1,t_2,\cdots,t_n)&=&
\bra{0} T\hat q(t_1)\hat q(t_2)\cdots\hat q(t_n)\ket{0}\\
&\equiv&{\bra{0,T}T\hat q(t_1)\hat q(t_2)\cdots\hat q(t_n)\ket{0,-T}
\over\bra{0,T}0,-T\rangle}\\
&=&{\int\DD q \,q(t_1)q(t_2)\cdots q(t_n)e^{iS}
\over\int\DD q \,e^{iS}},
\eeano
with no proportionality sign. The wave functions and exponential
factors in the numerator and denominator cancel.

To compute the numerator, we can once again use the trick we used in
perturbation theory in quantum mechanics, namely, adding a source to
the action. We define
\[
Z[J]={\int\DD q\,e^{i(S+\int dt\,J(t)q(t))}\over
\int\DD q\,e^{iS}}
={\bra{0}0\rangle_J\over\bra{0}0\rangle_{J=0}}.
\]
If we operate on $Z[J]$ with $i^{-1}\delta/\delta J(t_1)$, this gives
\beano
\left.\left({1\over i}
{\delta\over\delta J(t_1)}Z[J]\right)\right|_{J=0}
&=&\left.\left({\int\DD q\,q(t_1)e^{i(S+\int dt\,J(t)q(t))}
\over\int\DD q\,e^{iS}}
\right)\right|_{J=0}\\
&=&{\int\DD q \,q(t_1)e^{iS}
\over\int\DD q \,e^{iS}}\\
&=&{\bra{0,T}\hat q(t_1)\ket{0,-T}\over\bra{0,T}0,-T\rangle}
=\bra{0}\hat q(t_1)\ket{0}
\eeano
(The expectation values are evaluated in the {\em absence} of $J$.)

Repeating this procedure, we obtain a PI with several $q$'s in the
numerator. This ordinary product of $q$'s in the PI
corresponds, as discussed earlier in this section, to
a time-ordered product in the matrix element. So we make the following
conclusion:
\[
\left.\left({1\over i}{\delta\over\delta J(t_1)}\cdots
{1\over i}{\delta\over\delta J(t_n)}Z[J]\right)\right|_{J=0}
={\int\DD q \,q(t_1)\cdots q(t_n)e^{iS}
\over\int\DD q \,e^{iS}}\\
=\bra{0}T\hat q(t_1)\cdots\hat q(t_1)\ket{0}.\\
\]
For obvious reasons,
the functional $Z[J]$ is called the {\em generating functional} for
Green's functions; it is a very handy tool in quantum field
theory and in statistical mechanics.

How do we calculate $Z[J]$? Let us examine the numerator:
\[
N\equiv\int\DD q\,e^{i(S+\int dt\,J(t)q(t))}.
\]
Suppose initially that $S$ is the harmonic oscillator action (denoted
$S_0$):
\[
S_0=\int dt\,\left(\half m{\dot q}^2-\half m\om^2q^2\right),
\]
Then the corresponding numerator, $N_0$, is the non-Euclidean (\ie
real-time) version of the propagator $K_E^0[J]$ we used in Section
6. We can calculate $N_0[J]$ in the same way as $K_E^0[J]$. Since the
calculation repeats much of that of $K_E^0[J]$, we will be succinct.

By definition,
\[
N_0=\int \DD q(t)\,\exp{i\int dt
\left(\half m{\dot q}^2-\half m\om^2q^2+Jq\right)}.
\]
We do the path integral over a new variable $y$, defined by
$q(t)=q_c(t)+y(t)$, where $q_c$ is the classical solution. Then the PI
over $y$ is a constant (independent of $J$)
and we can avoid calculating it. (It will
cancel against the denominator in $Z[J]$.) Calling it $C$, we have
\[
N_0=C e^{iS_{0J}[q_c]},
\]
where
\[
S_{0J}[q_c]=\int dt
\left(\half m{\dot q_c}^2-\half m\om^2q_c^2+Jq_c\right)
=\half\int dt J(t)q_c(t),
\]
using the fact that $q_c$ satisfies the equation of
motion. We can write the classical path in terms of the Green's
function (to be determined shortly), defined by
\beq
\left({d^2\over dt^2}+\om^2\right)G(t,t')
=-i\delta(t-t').
\label{gf1}
\eeq
Then
\[
q_c(t)=-i\int dt'G(t,t')J(t').
\]
We can now write
\[
N_0=C \exp{\half\int dt dt'\,J(t)G(t,t')J(t')}.
\]
Dividing by the denominator merely cancels the factor $C$, giving our
final result:
\[
\fbox{$\displaystyle
Z[J]=\exp{\half\int dt dt'\,J(t)G(t,t')J(t')}.
$}
\]

We can solve \eqref{gf1} for the Green's function by going into
momentum space; the result is
\[
G(t,t')=G(t-t')=\int{dk\over2\pi}{i\over k^2-\om^2}
e^{-ik(t-t')}.
\]
However, there are poles on the axis of integration. (This problem did
not arise in Euclidean space; see \eqref{gf}.) 
The Green's function is ambiguous until
we give it a ``pole prescription'', \ie a boundary condition. But
remember that our time $T$ has a small, negative imaginary part. We
require that $G$ go to zero as $T\to\infty$. The correct pole
prescription then turns out to be
\beq
\fbox{$\displaystyle
G(t-t')=\int{dk\over2\pi}{i\over k^2-\om^2+i\eps}
e^{-ik(t-t')}.
$}
\label{new1}
\eeq

We could at this point do a couple of practice calculations to get
used to this formalism. Examples would be to compute perturbatively
the generating functional for an action which has terms beyond
quadratic (for example, a $q^4$ term), or to compute some Green's
function in either the quadratic or quartic theory. But since these
objects aren't really useful in quantum mechanics, without further
delay we will go directly
to the case of interest: quantum field theory.

\newpage\thispagestyle{empty}
\section[Perturbation Theory II]
{Green's Functions in Quantum Field Theory}

It is easy to generalize the PI to many degrees of freedom; we have in
fact already done so in Section 4, where particles move in two or
three dimensions. It is simply a matter of adding a new index to
denote the different degrees of freedom (be they the different
coordinates of a single particle in more than one dimension or the
particle index for a system of many particles).

One of the most important examples of a system
with many degrees of freedom is a field
theory: $q(t)\to\phi(\bfx,t)=\phi(x)$. Not only is this a system of
{\em many} degrees of freedom, but one of a continuum of degrees of
freedom. The passage from a discrete to continuous system in path
integrals can be done in the same way as in ordinary classical
field theory: we
can discretize the field (modeling it by a set of masses and springs,
for instance), do the usual path integral manipulations on the
discrete system, and take the continuum limit at the end of the
calculation. The final result is a fairly obvious generalization of
the one-particle results, so I will not dwell on the mundane details
of discretization and subsequent taking of the continuum limit.

The analog of the quantum mechanical propagator is the transition
amplitude to go from one field configuration $\phi(\bfx)$ at $t=0$ to
another $\phi'(\bfx')$ at $t=T$:
\beq
K(\phi'(\bfx'),T;\phi(\bfx),0)=\int \DD\phi e^{iS[\phi]},
\label{qft1}
\eeq
where $S$ is the field action, for instance
\beq
S[\phi]=\int d^4x\left(\half(\partial_\mu\phi)^2-\half m^2\phi^2\right)
\label{fsf}
\eeq
for the
free scalar field. In \eqref{qft1}
the integral is over all field configurations $\phi(x)$ obeying the
stated initial and final conditions.

In field theory, we are not really interested in this object. Rather
(as mentioned earlier), we are interested in Green's functions. Most of
the work required to translate \eqref{qft1} into an expression for a
Green's function (generating functional of Green's functions,
more precisely) has already been done in the last section, so let us
study a couple of cases.

\subsection{Free scalar field.}
For the free scalar field, whose action is given by \eqref{fsf}, the
generating functional is
\[
Z_0[J]={\bra{0}0\rangle_J\over\bra{0}0\rangle_{J=0}}.
\]
Both numerator and denominator
can be written in terms of PIs. The numerator is
\[
N_0=\int\DD \phi\, \exp{i\int d^4x
\left(\half(\partial_\mu\phi)^2-\half m^2\phi^2+J\phi\right)}.
\]
We write $\phi=\phi_c+\varphi$, where $\phi_c$ is the classical field
configuration, and integrate over the deviation from $\phi_c$. The
action can be written
\[
S[\phi_c+\varphi]=
\int d^4x\left(\half(\partial_\mu\phi_c)^2
-\half m^2\phi_c^2+J\phi_c\right)
+\int d^4x\left(\half(\partial_\mu\varphi)^2-\half m^2\varphi^2\right),
\]
where as usual there is no term linear in $\varphi$ since $\phi_c$ by
definition extremizes the classical action. So
\[
N_0=C \exp{i\int d^4x\left(\half(\partial_\mu\phi_c)^2
-\half m^2\phi_c^2+J\phi_c\right)},
\]
where
\[
C=\int\DD\varphi \exp{i\int d^4x\left(\half(\partial_\mu\varphi)^2
-\half m^2\varphi^2\right)}.
\]
$C$ is independent of $J$ and will cancel in $Z$. (Indeed, the
denominator is {\em equal to} $C$.)

Using the fact that $\phi_c$ obeys the classical equation
\[
(\partial^2+m^2)\phi_c=J,
\]
we can write
\[
N_0=C \exp{{i\over2}\int d^4x\,J(x)\phi_c(x)}.
\]
Finally, we can write $\phi_c$ in terms of the Klein-Gordon Green's
function, defined by
\[
(\partial^2+m^2)\Delta_F(x,x')=-i\delta^4(x-x').
\]
It is
\[
\phi_c(x)=i\int d^4x\,\Delta_F(x,x')J(x'),
\]
so
\[
\fbox{$\displaystyle
Z_0={N_0\over C}=
\exp{-\half\int d^4x d^4x'\,J(x)\Delta_F(x,x')J(x')}.
$}
\]
The Green's function is found by solving its equation in 4-momentum
space; the result is
\[
\Delta_F(x,x')=\int{d^4k\over(2\pi)^4}{i\over k^2-m^2+i\eps}
e^{-ik\cdot(x-x')}=\Delta_F(x-x'),
\]
adopting the same pole prescription as in \eqref{new1}.
Note that $\Delta_F$ is an even function,
$\Delta_F(x-x')=\Delta_F(x'-x)$.

Let us calculate a couple of Green's functions. These calculations are
reminiscent of those at the end of Section 6. As a first example,
consider
\[
G_0^{(2)}(x_1,x_2)=\bra{0}T\hat\phi(x_1)\hat\phi(x_2)\ket{0}
={1\over i^2}\left.\left({\delta^2\over\delta J(x_1)\delta J(x_2)}
Z_0[J]\right)\right|_{J=0}.
\]
Expanding $Z_0$ in powers of $J$,
\[
Z_0[J]=1-\half\int d^4x d^4x'\,
J(x)\Delta_F(x-x')J(x')+o(J^4).
\]
The term {\em quadratic} in $J$ is the only one that survives both
differentiation (which kills the ``1'') and the setting of $J$ to zero
(which kills all higher-order terms). So
\[
G_0^{(2)}(x_1,x_2)={1\over i^2}
{\delta^2\over\delta J(x_1)\delta J(x_2)}
\left(-\half\int d^4x d^4x'\,J(x)\Delta_F(x-x')J(x')\right).
\]
There arise two identical terms, depending on which derivative acts on
which $J$. The result is
\[
G_0^{(2)}(x_1,x_2)=\Delta_F(x_1-x_2).
\]
So the Green's function (or two-point function) in the quantum field
theory sense is also the Green's function in the usual
differential-equations sense.

As a second example, the four-point Green's function is
\[
G_0^{(4)}(x_1,x_2,x_3,x_4)={1\over i^4}\left.\left(
{\delta\over\delta J(x_1)}\cdots{\delta\over\delta J(x_4)}
\exp{-\half\int d^4x d^4x'\,J(x)\Delta_F(x,x')J(x')}
\right)\right|_{J=0}.
\]
This time the only part of the exponential that contributes is the
term with four $J$'s.
\[
G_0^{(4)}(x_1,x_2,x_3,x_4)=
{\delta\over\delta J(x_1)}\cdots{\delta\over\delta J(x_4)}
\half\left(-\half\int d^4x d^4x'\,J(x)\Delta_F(x,x')J(x')\right)^2.
\]
There are $4!=24$ terms, corresponding to the number of ways of
associating the derivatives with the $J$'s.
In 8 of them, the Green's functions which
arise are $\Delta_F(x_1-x_2)\Delta_F(x_3-x_4)$, and so on. 
The result is
\bea
G_0^{(4)}(x_1,x_2,x_3,x_4)&=&\Delta_F(x_1-x_2)\Delta_F(x_3-x_4)+
\Delta_F(x_1-x_3)\Delta_F(x_2-x_4)\nonumber\\
&&\qquad+\Delta_F(x_1-x_4)\Delta_F(x_2-x_3),
\label{g4free}
\eea
which can be represented diagramatically as in Figure 13.

\begin{figure}[hb]
\epsfysize=3cm
\centerline{\epsfbox{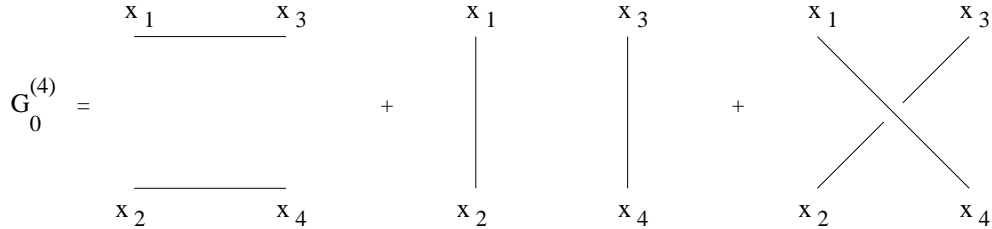}}
\caption{Diagrammatic representation of \eqref{g4free}. Each line
counts as a factor $\Delta_F$ with argument corresponding to the
endpoints of the line.}
\end{figure}

\subsection{Interacting scalar field theory.}
Usually, if the Lagrangian has a term beyond quadratic we can no longer
evaluate exactly the functional integral, and we must resort to
perturbation theory. The generating functional method is tailor-made
to do this in a systematic fashion. To be
specific, consider $\phi^4$ theory, defined by the Lagrangian density
\[
\LL=\half(\partial_\mu\phi)^2-\half m^2\phi^2-{\la\over4!}\phi^4.
\]
Then the generating functional is (up to an unimportant constant: we
will normalize ultimately so that $Z[J=0]=1$)
\[
Z[J]=C\int\DD\phi\,\exp{i\int d^4x\left(
\half(\partial_\mu\phi)^2-\half m^2\phi^2-{\la\over4!}\phi^4
+J\phi\right)}.
\]
Because of the quartic term, we cannot evaluate the functional
integral exactly. But we can use a trick we first saw when discussing
perturbation theory in quantum mechanics: replacing the higher-order
term by a functional derivative with respect to $J$:
\beano
Z[J]&=&C\int\DD\phi\,\exp\left\{{-i{\la\over4!}\int d^4x\,\phi^4}
\right\}
\exp{i\int d^4x\left(
\half(\partial_\mu\phi)^2-\half m^2\phi^2+J\phi\right)}\\
&=&C\int\DD\phi\,\exp\left\{{-i{\la\over4!}\int d^4x\
\left({1\over i}{\delta\over\delta J(x)}\right)^4}\right\}
\exp{i\int d^4x\left(
\half(\partial_\mu\phi)^2-\half m^2\phi^2+J\phi\right)}.
\eeano
We can pull the first exponential out of the integral; the functional
integral which remains is that for $Z_0$. Adjusting the constant $C$
so that $Z[J=0]=0$, we get
\[
Z[J]={
  \exp\left\{{-i{\la\over4!}\int d^4x
  \left({1\over i}{\delta\over\delta J(x)}\right)^4}\right\}
  \exp{-\half\int d^4x d^4x'\,J(x)\Delta_F(x,x')J(x')}
\over
  \left.\left(\exp\left\{{-i{\la\over4!}\int d^4x
  \left({1\over i}{\delta\over\delta J(x)}\right)^4}\right\}
  \exp{-\half\int d^4x d^4x'\,J(x)\Delta_F(x,x')J(x')}
  \right)\right|_{J=0}
}.
\]
This expression now enables us to compute a perturbative expansion for
any Green's function we desire. This is a rather mechanical job, and
the only way to learn it is by doing lots of examples. To illustrate
the method, let us look at $G^{(2)}(x_1,x_2)$ to the first nontrivial
order in $\la$.

We have
\[
G^{(2)}(x_1,x_2)={
  \left.\left({1\over i^2}{\delta^2\over\delta J(x_1)\delta J(x_2)}
  \exp\left\{{-i{\la\over4!}\int d^4x
  \left({1\over i}{\delta\over\delta J(x)}\right)^4}\right\}
  \exp{-\half\vev{J_a\Delta_{Fab}J_b}}
  \right)\right|_{J=0}
\over
  \left.\left(\exp\left\{{-i{\la\over4!}\int d^4x
  \left({1\over i}{\delta\over\delta J(x)}\right)^4}\right\}
  \exp{-\half\vev{J_a\Delta_{Fab}J_b}}
  \right)\right|_{J=0}
},
\]
where as in Section 6
$\vev{\cdots}$ implies integration over the positions of
the $J$'s. In both numerator and denominator, we can expand both
exponentials. The only terms that survive are those that have the same
total number of derivatives and $J$'s. Let us look at the term linear
in $\la$ in the numerator. There are six derivatives, so we need the
term from the expansion of the second exponential with six $J$'s. For
this term, we get the following expression for the numerator:
\[
-{\delta^2\over\delta J(x_1)\delta J(x_2)}\left(-i{\la\over4!}\right)
\int d^4x\left({\delta\over\delta J(x)}\right)^4
{1\over3!}\left(-\half\right)^3
\vev{J_a\Delta_{Fab}J_b}\vev{J_c\Delta_{Fcd}J_d}
\vev{J_e\Delta_{Fef}J_f}.
\]
There are now a total of $6!=720$ terms! However, only two distinct
analytical expressions result. The first of these
arises if the derivatives at $x_1$ and $x_2$ act on different
$\vev{\cdots}$'s. A little combinatorial head-scratching tells us that
there are 576 such terms, yielding the following expression:
\beq
-{i\la\over2}\int d^4x\,\Delta_F(x_1-x)\Delta_F(x-x)\Delta_F(x-x_2),
\label{new2}
\eeq
which can be represented pictorially as in Figure 14 (a).

The only other expression arises when
the derivatives at $x_1$ and $x_2$ act on the same
$\vev{\cdots}$. This accounts for the remaining 144 terms; the analytic
form which results is
\beq
-{i\la\over8}\Delta_F(x_1-x_2)\int d^4x\,\Delta_F(x-x)^2,
\label{new2a}
\eeq
corresponding to the diagram in Figure 14(b).

\begin{figure}[ht]
\epsfysize=3cm
\centerline{\epsfbox{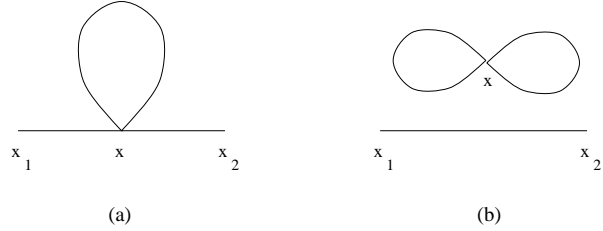}}
\caption{Diagrammatic representation of (a) \eqref{new2} and (b)
  \eqref{new2a}. Intersection points represent vertices, and count as
  a factor $-i\la\int d^4x$. Symmetry factors complete the
  association of an analytic expression with each diagram.}
\end{figure}

The denominator can be evaluated in a similar fashion; the Green's
function to order $\la$ is
\[
G^{(2)}(x_1,x_2)={
\begin{array}{c}
\Big\{\Delta_F(x_1-x_2)
-{i\la\over2}\int d^4x\,\Delta_F(x_1-x)\Delta_F(x-x)\Delta_F(x-x_2)\\
\qquad-{i\la\over8}\Delta_F(x_1-x_2)\int d^4x\,\Delta_F(x-x)^2
+o(\la^2)\Big\}
\end{array}
\over
1-{i\la\over8}\int d^4x\,\Delta_F(x-x)^2+o(\la^2)
}.
\]
Since we have only computed the numerator and denominator to order
$\la$, we can rewrite this expression in the following way:
\[
G^{(2)}(x_1,x_2)={
\left(\begin{array}{c}
\left\{\Delta_F(x_1-x_2)
-{i\la\over2}\int d^4x\,\Delta_F(x_1-x)\Delta_F(x-x)\Delta_F(x-x_2)
+o(\la^2)\right\}\\
\qquad\qquad
\times\left\{1-{i\la\over8}\int d^4x\,\Delta_F(x-x)^2+o(\la^2)\right\}
\end{array}\right)
\over
1-{i\la\over8}\int d^4x\,\Delta_F(x-x)^2+o(\la^2)
}.
\]
We can now cancel the second factor in the numerator against
the denominator, {\em to order $\la$}, resulting in
\[
G^{(2)}(x_1,x_2)=\Delta_F(x_1-x_2)
-{i\la\over2}\int d^4x\,\Delta_F(x_1-x)\Delta_F(x-x)\Delta_F(x-x_2)
+o(\la^2).
\]
This factorization of the numerator into a part containing no factors
independent of the external position times the denominator occurs to
all orders, as can be proven fairly cleanly via a combinatoric
argument. The conclusion is that so-called disconnected parts (parts
of diagrams not connected to any external line) cancel
from Green's functions, a fact which simplifies greatly the
calculation of these objects.

It cannot be overemphasized
that there are only three ways to get accustomed to this formalism:
practice, practice,
and practice. Other reasonable exercises are the calculation of
$G^{(2)}$ to order $\la^2$ and the calculation of $G^{(4)}$ to order
$\la^2$. $\phi^3$ theory is also a useful testing ground for the
techniques discussed in this section.

\newpage\thispagestyle{empty}
\section[Instantons]
{Instantons in Quantum Mechanics}

\subsection{General discussion}
It has already been briefly mentioned that in quantum mechanics 
certain aspects of a problem can be overlooked in a perturbative
treatment. One example occurs if we have a harmonic oscillator with a
cubic anharmonic term: $V(q)=\half m\om^2 q^2+\la q^3$ (Figure 15).

\begin{figure}[hb]
\epsfysize=5cm
\centerline{\epsfbox{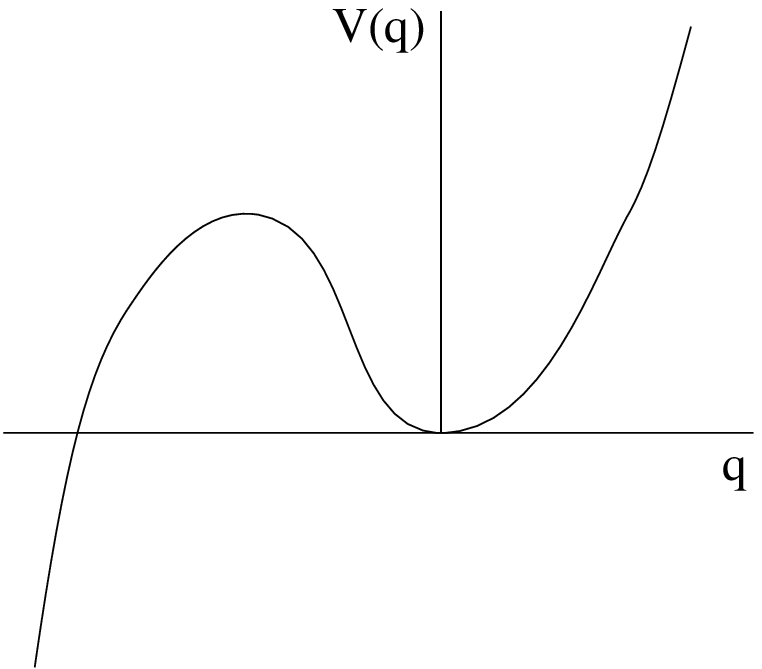}}
\caption{$V(q)=\half m\om^2 q^2+\la q^3$.}
\end{figure}

We can calculate corrections to harmonic oscillator wave functions
and energies perturbatively in $\la$, to any desired
order, blissfully ignorant of a serious pathology in
the model.
As can be seen from Figure 15, this model has no ground state: the
potential energy is unbounded as $q\to-\infty$, a point completely
invisible to perturbation theory.

A second example is the double-well potential,
$V(q)={\la\over4!}(q^2-a^2)^2$ (Figure 16).
\begin{figure}[hb]
\epsfysize=4.5cm
\centerline{\epsfbox{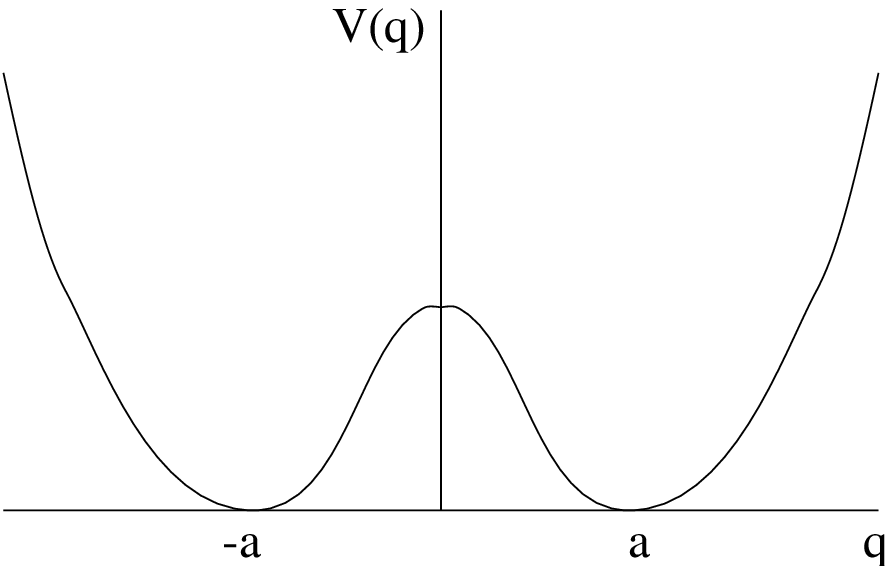}}
\caption{$V(q)={\la\over4!}(q^2-a^2)^2$.}
\end{figure}
There are two classical ground states. We can ignore this fact and
expand $V$ about one of the minima; it then takes the form of
a harmonic oscillator about that minimum plus anharmonic terms
(both cubic and quartic). We can then compute perturbative corrections
to the wave functions and energies, and never see any evidence of the
other minimum. Were we to expand about the other minimum, we would
produce an identical set of perturbative corrections. By symmetry the
ground state energies calculated perturbatively to {\em any order} will
be the same for the expansions about the two minima, so it appears
that we have degenerate ground states. But in fact the ground state is
not degenerate: a  nonperturbative energy splitting separates the true
ground state (an even function of $q$) from the first excited state
(an odd function); this splitting is not seen in perturbation theory.

We will examine this second example using PIs, the main goal being to
calculate the energy splitting between the two candidate ground states.

Let us first recall the PI expression for the Euclidean propagator:
\[
K_E(q',{\be\over2};q,-{\be\over2})=
\bra{q'}e^{-\be H/\hbar}\ket{q}=
\int\DD q\,e^{-S_E/\hbar},
\]
where 
\[
S_E=\int_{-\be/2}^{\be/2}d\tau\left(\half m{\dot q}^2+V(q)\right).
\]
Henceforth, we will set $m\to1$. $K_E$ is useful because we can write
it as
\beq
K_E=\sum_n\bra{q'}n\rangle\langle n\ket{q}e^{-\be E_n/\hbar};
\label{inst1}
\eeq
in the limit $\be\to\infty$, this term will be dominated by the
lowest-energy states. I say ``states'' here rather than ``state''
because we must calculate the two lowest-energy eigenvalues to get
the splitting of the (perturbatively degenerate) lowest-energy states
in the double-well potential.

We will evaluate the PI using an approximation known as the
semiclassical approximation, or alternatively as the method of steepest
descent. To illustrate it, consider the following integral
\[
I=\int_{-\infty}^\infty dx\,e^{-S(x)/\hbar},
\]
where $S(x)$ is a function with several local minima (Figure 17).
\begin{figure}[hb]
\epsfysize=5cm
\centerline{\epsfbox{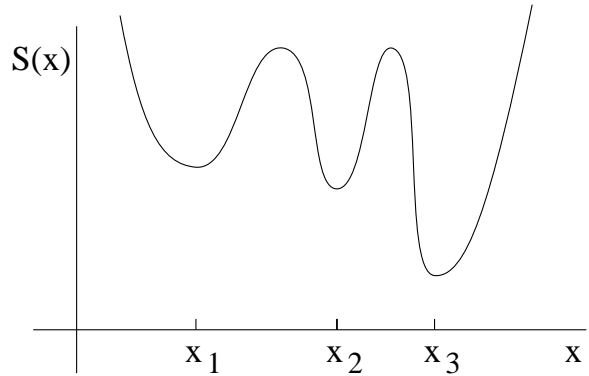}}
\caption{Potential with several minima.}
\end{figure}

Suppose we are interested in this integral as $\hbar\to0$. Then the
integral will be dominated by the minima of $S$; we can approximate it
by a series of Gaussian integrations, one for each minimum of
$S$. If $x_i$ is such a minimum, then in its vicinity $S(x)\simeq
S(x_i)+\half(x-x_i)^2 S''(x_i)$; we can write
\beq
I\simeq I_1+I_2+I_3+\cdots,
\label{inst2}
\eeq
where
\beano
I_i&=&\int_{-\infty}^\infty dx\,
\exp{-[S(x_i)+\half(x-x_i)^2 S''(x_i)]/\hbar}\\
&=&e^{-S(x_i)/\hbar}\sqrt{2\pi\hbar\over S''(x_i)}.
\eeano

Anharmonicities of $S$ appear as corrections of order $\hbar$ to
$I$. (This can be easily seen, for example, by considering a specific
case such as $S(x)=ax^2+bx^4$.)

We will compute the PI \eqref{inst1} in the semi-classical approximation,
where the analog of the
$x_i$ in the above example
will be classical paths (extremum of the action $S_E[q]$).

Suppose, then, that $q_c(\tau)$ is the classical solution to the
problem
\[
{d^2\over d\tau^2}q={\partial V(q)\over\partial q},\qquad
q(-\be/2)=q,\qquad q(\be/2)=q'.
\]
We can write $q(\tau)=q_c(\tau)+y(\tau)$; the action is
\bea
S_E[q_c+y]&=&\int_{-\be/2}^{\be/2} d\tau
\left(\half {{\dot q}_c+\dot y}^2+V(q_c+y)\right)\nonumber\\
&=&\int_{-\be/2}^{\be/2} d\tau
\left(\half {{\dot q}_c}^2+V(q_c)\right)+
(\mbox{linear in $y$})\nonumber\\
&&\qquad+\int_{-\be/2}^{\be/2} d\tau
\left(\half {\dot y}^2+\half V''(q_c)y^2\right)+\cdots.
\label{inst3}
\eea
The term linear in $y$ vanishes for the usual reason, and the higher
order terms not written down
are of cubic or higher order in
$y$. Neglecting these (which give order $\hbar$ corrections to the
PI), the propagator becomes
\[
K_E=\int\DD q\,e^{-S_E/\hbar}
=e^{-S_E[q_c]/\hbar}\int\DD y\,
\exp{-\int d\tau\left(\half {\dot y}^2+\half V''(q_c)y^2\right)/\hbar}.
\]
The functions $y(\tau)$ over which we integrate satisfy the boundary
conditions $y(-\be/2)=y(\be/2)=0$.
The PI, being Gaussian, can be done exactly; it is not as
straightforward as the harmonic oscillator PI since $V''(q_c)$ depends
on $\tau$. While we have often managed to {\em avoid} evaluating PIs,
here we must evaluate it. (Unfortunately, this is rather difficult.)

To this end, we can use a generalization of the Fourier expansion
technique mentioned in Section 2.2.2. We can rewrite the action as
\beq
S_E=\int d\tau\left(\half{\dot y}^2+\half V''(q_c)y^2\right)
=\half\int d\tau\, y\left(-{d^2\over d\tau^2}+V''(q_c)\right)y.
\label{inst3a}
\eeq
The Schroedinger-like equation
\[
\left(-{d^2\over d\tau^2}+V''(q_c)\right)y=\la y,
\qquad y(-\be/2)=y(\be/2)=0
\]
has a complete, orthonormal set of solutions; let the solutions and
eigenvalues be $y_k(\tau)$ and $\la_k$, respectively. The
orthonormality relation is
\[
\int_{-\be/2}^{\be/2} d\tau y_k(\tau)y_l(\tau)=\delta_{kl}.
\]
Then we can substitute $y(\tau)=\sum_k a_k y_k(\tau)$ in
\eqref{inst3a}, giving
\[
S_E=\half\int d\tau\sum_k a_k y_k
\left(-{d^2\over d\tau^2}+V''(q_c)\right)\sum_l a_l y_l
=\half\sum_{k,l}a_k a_l \la_l \int d\tau y_k y_l
=\half\sum_k a_k^2\la_k.
\]
The PI can now be written as an integral over all possible values of
the coefficients $\{a_k\}$. This gives
\beq
K_E=J'\int\prod_k da_k\,e^{-\sum_k a_k^2\la_k/2\hbar},
\label{inst5}
\eeq
where $J'$ is the Jacobian of the transformation from $y(\tau)$ to
$\{a_k\}$. \eqref{inst5} is a product of uncoupled Gaussian integrals;
the result is
\[
K_E=J'\prod_k\left({2\pi\hbar\over\la_k}\right)^{1/2}
=J'\prod_k(2\pi\hbar)^{1/2}(\prod_k \la_k)^{-1/2}
=J'\prod_k(2\pi\hbar)^{1/2}
{\det}^{-1/2}\left(-{d^2\over d\tau^2}+V''(q_c)\right),
\]
where we have written the product of eigenvalues as the determinant of
the Schroedinger operator on the space of functions vanishing at
$\pm\be/2$.

We can write $J=J'\prod_k(2\pi\hbar)^{1/2}$, giving
\[
K_E=J{\det}^{-1/2}\left(-{d^2\over d\tau^2}+V''(q_c)\right)
(1+o(\hbar)),
\]
where the $o(\hbar)$ corrections can in principle be computed from the
neglected beyond-quadratic terms in \eqref{inst3}. We will not be
concerned with these corrections, and henceforth we will drop the
$(1+o(\hbar))$.

\subsection{Single Well in the Semiclassical Approximation}
Before looking at the double well, it is worthwhile examining the
single well, defined be
\[
V(q)=\half\om^2q^2+{\la\over4!}q^4.
\]
The classical equation is
\[
{d^2\over d\tau^2}q=V'(q).
\]
Note that this is the equation of motion for a particle moving in a
potential $-V(q)$. If we choose the initial and final points $q=q'=0$,
then the classical solution is simply $q_c(\tau)=0$; furthermore,
$V''(q_c)=V''(0)=\om^2$, and
\[
K_E=J{\det}^{-1/2}\left(-{d^2\over d\tau^2}+\om^2\right).
\]
The evaluation of the determinant is not terribly difficult (the
eigenvalues can be easily found; their product can be found in a table
of mathematical identities); the result, for large $\be$, is
\[
K_E=\left({\om\over\pi\hbar}\right)^{1/2}e^{-\be\om/2}.
\]
From \eqref{inst1}, we can extract the ground state energy since, for
large $\be$, $K_E\sim\exp-E_0\be/\hbar$. We find $E_0=\hbar\om/2$ up to
corrections of order $\la\hbar^2$. We have discovered an incredibly
complicated way of calculating the ground state energy of the harmonic
oscillator!

\subsection{Instantons in the Double Well Potential}
Let us now study a problem of much greater
interest: the double well. We will see
that configurations known as ``instantons'' make a non-perturbative
correction to the energies. We wish to evaluate the PI
\[
K_E=\int_{q,-\be/2}^{q',\be/2}\DD q\,e^{-S_E},
\]
where
\[
S_E=\int d\tau\left(\half{\dot q}^2+{\la\over4!}(q^2-a^2)^2\right),
\]
for $\be\to\infty$. As explained above, the PI is dominated by minima
of $S_E$, \ie by classical solutions. The classical equation
corresponds to a particle moving in the potential $-V(q)$
(Figure 18); the ``energy'' $E=\half{\dot q}^2-V(q)$ is conserved.

\begin{figure}[hb]
\epsfysize=5cm
\centerline{\epsfbox{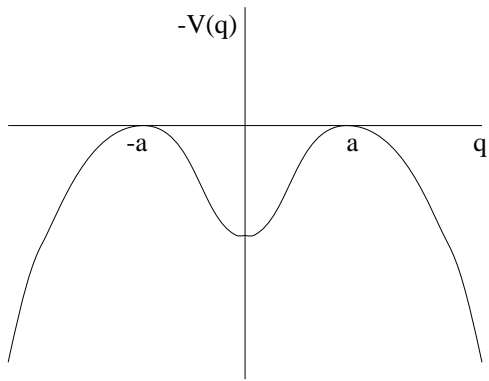}}
\caption{Inverted double-well potential.}
\end{figure}

Let us examine classical solutions, taking the boundary values $q,q'$
of the classical solution corresponding to the maxima of $-V$, $\pm
a$. In the limit $\be\to\infty$, these will be solutions of zero
``energy'', since as $\tau\to\pm\infty$ both the kinetic and potential
``energy'' vanish.

First, if $q=q'=a$ (an identical argument applies if $q=q'=-a$), the
obvious classical
solution is $q(\tau)=a$; a quadratic approximation about this
constant solution would be identical to the single-well case discussed
above.

But what if $q=-a$ and $q'=a$ (or vice-versa)? Then the obvious
classical solution corresponds to the particle initially sitting atop the
maximum of $-V$ at $-a$, rolling towards $q=0$ after a very long
(infinite, in the limit $\be\to\infty$) time, and ending up at
rest at the other maximum of $-V$ as $\tau\to\infty$ (Figure 19).
\begin{figure}[hb]
\epsfysize=5cm
\centerline{\epsfbox{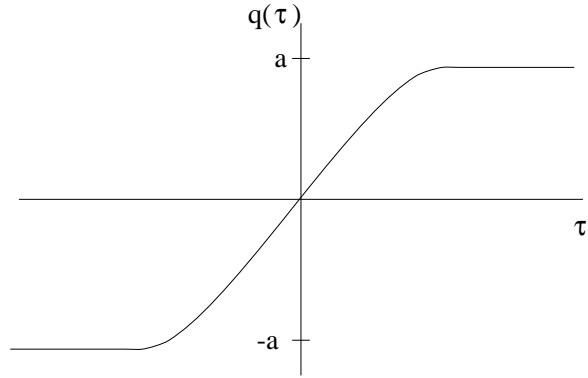}}
\caption{Instanton in the double-well potential.}
\end{figure}

We can get the analytical form of this solution: setting $E\to0$, we
have
\[
\half{\dot q}^2=V(q),\qquad\mbox{or}\qquad
{dq\over d\tau}=\pm\sqrt{\la\over12}(q^2-a^2).
\]
There are a family of solutions interpolating between $-a$ and $a$:
\beq
q(\tau)=a\tanh{\om\over2}(\tau-\tau_0),
\label{instanton}
\eeq
where $\om=\sqrt{\la a^2/3}$ and where $\tau_0$ is an integration
constant which corresponds to the time at which the solution crosses
$q=0$.

This solution is much like a topological soliton in field theory,
except that it is localized in time rather than in space.
One could argue that the solution doesn't
appear to be localized: $q$ goes to different values as
$\tau\to\pm\infty$. But these are just different, but physically
equivalent, ground states, so we can say that the instanton is a
configuration which interpolates between two ground states; the
system is in a ground state except for a brief time -- an
``instant''. For this reason, the solution is known
as an {\em instanton}.

I called the two solutions $q(\tau)=a$ and
$q(\tau)=a\tanh{\om\over2}(\tau-\tau_0)$ the {\em obvious} classical
solutions because there are an infinite number of approximate
classical solutions which are potentially important in the PI. Since
the instanton is localized in time, and since the total time interval
$\be$ is very large (in particular,
much larger than the instanton width), a series of widely-separated
instantons and anti-instantons (configurations interpolating between
$+a$ and $-a$) is also a solution, up to exponentially small
interactions between neighbouring instantons and anti-instantons. Such
a configuration is shown in Figure 20, where the horizontal scale has
been determined by the duration of imaginary time $\be$; on this
scale the instanton and anti-instanton appear as step
functions.

\begin{figure}[hb]
\epsfysize=5cm
\centerline{\epsfbox{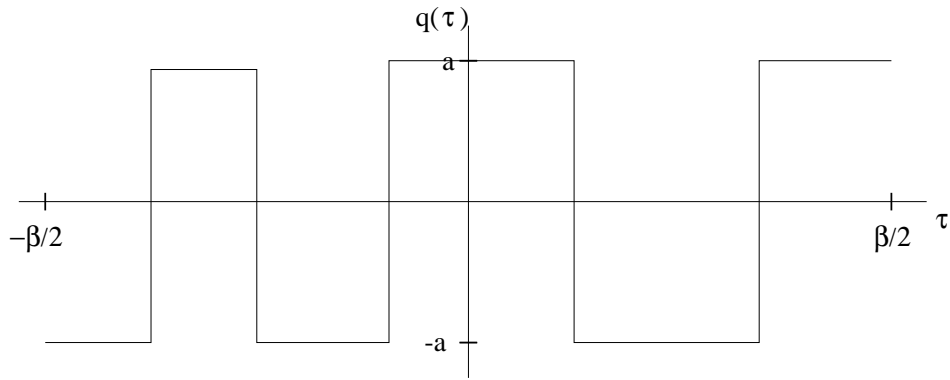}}
\caption{Multi-instanton configuration.}
\end{figure}

It is clear than an instanton must be followed by an anti-instanton,
and that if the asymptotic values of the position are $+a$ and $+a$
the classical solution must contain anti-instanton-instanton pairs
whereas if they are $-a$ and $+a$ we need an extra instanton at the
beginning.

Let us choose first limiting values $q(-\be/2)=q(+\be/2)=+a$. Then we
are interested in
\[
K_E=\int_{a,-\be/2}^{a,\be/2}\DD q\,e^{-S_E}.
\]
In the spirit of \eqref{inst2}, in the steepest-descent approximation
$K_E$ is equal to the sum of PIs evaluated about
{\em all} classical solutions. The classical solutions are:
$q_c(\tau)=a$;
$q_c=$anti-instanton-instanton$\equiv AI$; $q_c=AIAI$; \etcc where the
positions of the $A$s and $I$s are not determined, and must be
integrated over. Schematically, we may write
\beq
K_E=K_E^0+K_E^2+K_E^4+\cdots,
\label{new3}
\eeq
where the superscript denotes the
total number of $I$s or $A$s. Let us discuss the first couple of
contributions in some detail.

\noindent $q_c=a$: This case is essentially equivalent to the
single-well case discussed above, and we get
\[
K_E^0=\sqrt{\om\over\pi\hbar}e^{-\be\om/2},
\]
where $\om=(\la a^2/3)^{1/2}$ is the frequency of small oscillations
about the minimum of $V$.

\noindent $q_c=AI$: This case is rather more interesting (that is to
say, complicated!). Let us suppose that the classical solution around
which we expand consists of an anti-instanton at time $\tau_1$ and an
instanton at $\tau_2$ (Figure 21); clearly $\tau_2>\tau_1$.

\begin{figure}[hb]
\epsfysize=5cm
\centerline{\epsfbox{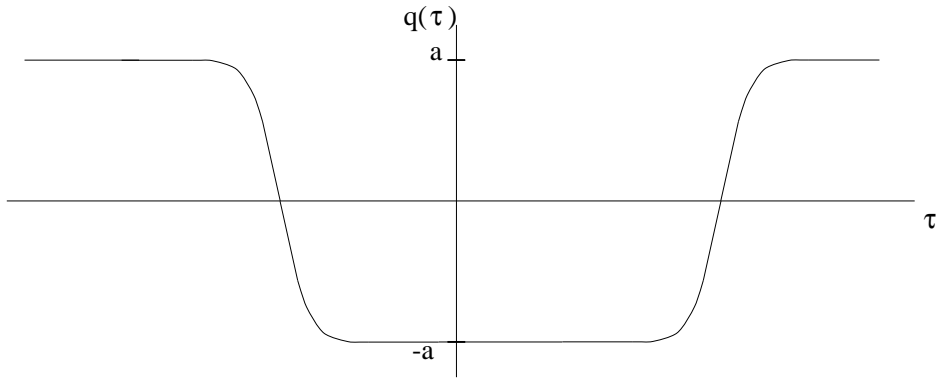}}
\caption{Anti-instanton-instanton.}
\end{figure}

Then we can write $q=q_c+y$, and
\[
S_E[q]=S_E[q_c]+S_E^{\mathrm{quad}}[y].
\]
We can evaluate $S_E[q_c]$: it is twice the action of a single
instanton (assuming the $I$ and $A$ are sufficiently far apart that any
interaction is negligible): $S_E[q_c]=2S_E^{\mathrm{inst}}$. The
one-instanton action $S_E^{\mathrm{inst}}$ is
\[
S_E^{\mathrm{inst}}
=\left.\int d\tau\left(\half{\dot q}^2+V(q)\right)
\right|_{\mathrm{inst}}=2\left.
\int d\tau V(q)\right|_{\mathrm{inst}}.
\]
With the instanton profile given by \eqref{instanton}, the result is
\[
S_E^{\mathrm{inst}}=\sqrt{\la\over3}{2a^3\over3}.
\]

To evaluate the PI with the action $S_E^{\mathrm{quad}}[y]$, let us
divide the imaginary time interval into two semi-infinite regions $I$
and $II$, where the boundary between the two regions is between and well
away from the $A$ and the $I$ (Figure 22).
\begin{figure}[hb]
\epsfysize=5cm
\centerline{\epsfbox{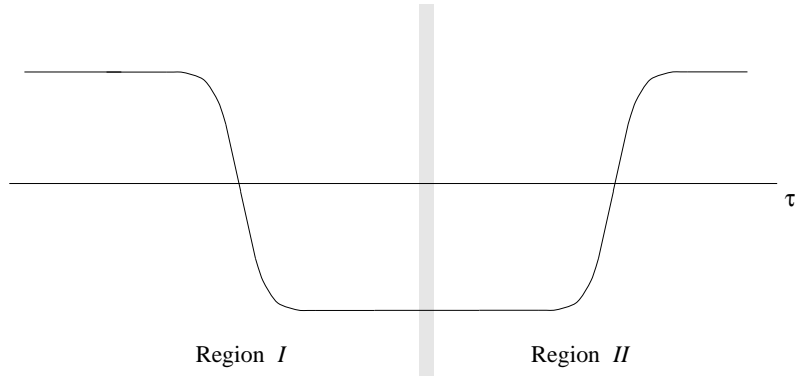}}
\caption{Division of imaginary time into two regions, one containing
  the anti-instanton, the other containing the instanton.}
\end{figure}

Then we can write
\beq
K_E^2={\be^2\over2}e^{-2S_E^{\mathrm{inst}}}
\int_{I+II}\DD y\,e^{-S_E^{\mathrm{quad}}/\hbar}.
\label{inst6}
\eeq
Here the first factor represents integration over the positions of the
$A$ and $I$ (remember that the $A$ must be to the left of the $I$!).
The quadratic action can be written
\[
S_E^{\mathrm{quad}}={S_E^{\mathrm{quad}}}_{I}+{S_E^{\mathrm{quad}}}_{II},
\]
where ${S_E^{\mathrm{quad}}}_{I}$ is the quadratic action in the presence
of an anti-instanton
and ${S_E^{\mathrm{quad}}}_{II}$ is that in the presence of an
instanton. 

Then the PI separates into two factors:
\beq
\int_{I+II}\DD y\,e^{-S_E^{\mathrm{quad}}/\hbar}
=\int_{I}\DD y\,e^{{-S_E^{\mathrm{quad}}}_{I}/\hbar}
\cdot\int_{II}\DD y\,e^{{-S_E^{\mathrm{quad}}}_{II}/\hbar},
\label{inst6a}
\eeq
where there is an implied integration over the intermediate position
at the boundary of the two regions.
The quadratic no-instanton PI also separates into two factors:
\beq
\int\DD y\,e^{-S_E^{\mathrm{quad},0}/\hbar}
=\int_{I}\DD y\,e^{{-S_E^{\mathrm{quad},0}}_{I}/\hbar}
\times\int_{II}\DD y\,e^{{-S_E^{\mathrm{quad},0}}_{II}/\hbar},
\label{inst7}
\eeq
where the superscript ``0'' denotes that this is the PI about a
no-instanton (constant) background.
We can combine \eqref{inst6a} and \eqref{inst7} to give:
\beq
\int_{I+II}\DD y\,e^{-S_E^{\mathrm{quad}}/\hbar}=
\int\DD y\,e^{-S_E^{\mathrm{quad},0}/\hbar}
{
\int_{I}\DD y\,e^{{-S_E^{\mathrm{quad}}}_{I}/\hbar}
\over
\int_{I}\DD y\,e^{{-S_E^{\mathrm{quad},0}}_{I}/\hbar}
}
{
\int_{II}\DD y\,e^{{-S_E^{\mathrm{quad}}}_{II}/\hbar}
\over
\int_{II}\DD y\,e^{{-S_E^{\mathrm{quad},0}}_{II}/\hbar}
}.
\label{inst8}
\eeq
But
\beq
{
\int_{I}\DD y\,e^{{-S_E^{\mathrm{quad}}}_{I}/\hbar}
\over
\int_{I}\DD y\,e^{{-S_E^{\mathrm{quad},0}}_{I}/\hbar}
}
=
{
\int\DD y\,e^{{-S_E^{\mathrm{quad}}}/\hbar}
\over
\int\DD y\,e^{{-S_E^{\mathrm{quad},0}}/\hbar}
}
\label{inst9}
\eeq
and similarly for the last factor in \eqref{inst8}, so we obtain
\[
\int_{I+II}\DD y\,e^{-S_E^{\mathrm{quad}}/\hbar}=
\sqrt{\om\over\pi\hbar}e^{-\be\om/2}R^2,
\]
where $R$ is the ratio of the PI in the presence and absence of an
instanton (or, equivalently, anti-instanton) given in \eqref{inst9}.
Substituting this into \eqref{inst6},
\[
K_E^2=e^{-2S_E^{\mathrm{inst}}/\hbar}\sqrt{\om\over\pi\hbar}
e^{-\be\om/2}R^2{\be^2\over2}.
\]
A similar argument gives
\[
K_E^4=e^{-4S_E^{\mathrm{inst}}/\hbar}\sqrt{\om\over\pi\hbar}
e^{-\be\om/2}R^4{\be^4\over4!},
\]
and so on for subsequent terms in the expansion \eqref{new3}.

Summing these contributions, we get
\beano
K_E&=&\sqrt{\om\over\pi\hbar}e^{-\be\om/2}\left(
1+{\left(\be R e^{-S_E^{\mathrm{inst}}/\hbar}\right)^2\over2!}
+{\left(\be R e^{-S_E^{\mathrm{inst}}/\hbar}\right)^4\over4!}+\cdots
\right)\\
&=&\sqrt{\om\over\pi\hbar}e^{-\be\om/2}
\cosh\left(\be R e^{-S_E^{\mathrm{inst}}/\hbar}\right)\\
&=&\half\sqrt{\om\over\pi\hbar}e^{-\be\om/2}
\left(e^{\be R e^{-S_E^{\mathrm{inst}}/\hbar}}+
e^{-\be R e^{-S_E^{\mathrm{inst}}/\hbar}}\right).
\eeano
Now we must recall why we're calculating this object in the first
place. The propagator can be written as in \eqref{inst1}:
\[
K_E=\sum_n\bra{a}n\rangle\langle n\ket{a}e^{-\be E_n/\hbar}.
\]
By comparing these two expressions we see that the lowest two energies
are
\[
{\hbar\om\over2}-\hbar R e^{-S_E^{\mathrm{inst}}/\hbar}
\qquad\mathrm{and}\qquad
{\hbar\om\over2}+\hbar R e^{-S_E^{\mathrm{inst}}/\hbar}.
\]
So the energy splitting is given by
\beq
\fbox{$\displaystyle
\Delta E=2\hbar R e^{-S_E^{\mathrm{inst}}/\hbar}.
$}
\label{esplit}
\eeq
$\Delta E$
is clearly non-perturbative: it cannot be expanded as a power
series in $\hbar$ (or, equivalently, in $\la$).

In principle, we should calculate the ratio
\[
R={(\mbox{instanton background PI})\over
(\mbox{constant background PI})}\sim
\mbox{ratio of determinants},
\]
but I don't know how to compute it other than by doing
a very arduous, technical calculation; luckily, time will not permit
it. The interested reader can consult the book by Sakita for a
discussion of this calculation.

As a final note, we have calculated the PI with $q=q'=a$; a good
exercise is to do the analogous calculation for $q=-a$, $q'=a$.

\subsection{Instantons in a Periodic Potential}

Consider a particle moving in a one-dimensional periodic potential
(Figure 23).
\begin{figure}[hb]
\epsfysize=4cm
\centerline{\epsfbox{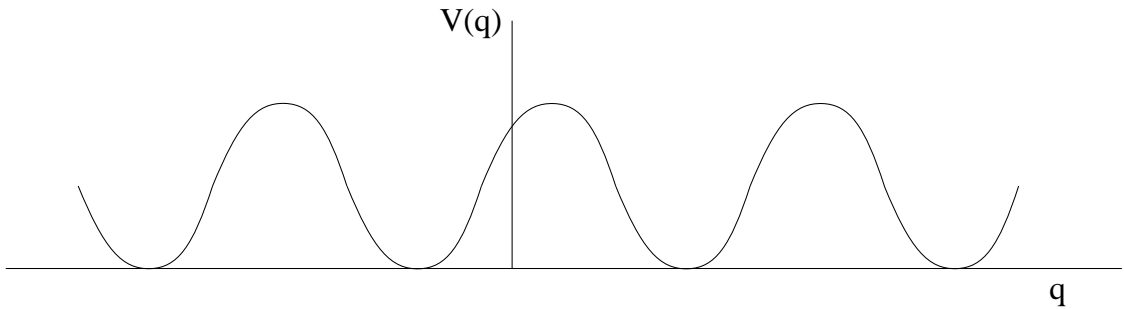}}
\caption{One-dimensional periodic potential.}
\end{figure}

With two minima, as we have seen,
instantons enable us to calculate the energy splitting between the
lowest-energy states of even and odd parity. In a periodic
potential, we
will see that a continuum of energies arise. 

Let us label the
classical minima of $V$ by an integer, $j$. 
Clearly this model will have solutions analogous to the instantons
above, going from any minimum of $V$ to the adjacent minimum.
We define an instanton as the classical
solution going from any $j$ to $j+1$, and an anti-instanton as that
going from $j$ to $j-1$. 
Then the Euclidean PI to go from
$j=0$ to $j=0$, for
instance, can be computed in a manner similar to the calculation
of the previous section. This time any number and any order of
instantons and anti-instantons are possible, subject to the constraint
that $n_I=n_A$.

A calculation similar to that of the previous section results in the
following expression for the propagator:
\beano
K_E(0,\be/2;0,-\be/2)&=&\sqrt{\om\over\pi\hbar}e^{-\be\om/2}
\sum_{n=0}^\infty{1\over n!^2}\left(\underbrace{
e^{-S_E^{\mathrm{inst}}/\hbar}R\be}_{\equiv Q}\right)^{2n}\\
&=&\sqrt{\om\over\pi\hbar}e^{-\be\om/2}
\sum_{n,n'=0}^\infty{Q^n\over n!}{Q^{n'}\over n'!}
\int_0^{2\pi}{d\theta\over2\pi}e^{i\theta(n-n')}\\
&=&\sqrt{\om\over\pi\hbar}e^{-\be\om/2}
\int_0^{2\pi}{d\theta\over2\pi}
\sum_{n=0}^\infty{(Q e^{i\theta})^n\over n!}
\sum_{n'=0}^\infty{(Q e^{-i\theta})^{n'}\over n'!}\\
&=&\sqrt{\om\over\pi\hbar}e^{-\be\om/2}
\int_0^{2\pi}{d\theta\over2\pi}
\exp\{{Q e^{i\theta}}\}\exp\{{Q e^{-i\theta}}\}\\
&=&\sqrt{\om\over\pi\hbar}e^{-\be\om/2}
\int_0^{2\pi}{d\theta\over2\pi}
\exp\left\{{2\be R e^{-S_E^{\mathrm{inst}}/\hbar}\cos\theta}\right\}.
\eeano
The derivation of this is a worthwhile exercise.
From it, we can read the energies:
\[
E(\theta)={\hbar\om\over2}-2\hbar R
e^{-S_E^{\mathrm{inst}}/\hbar}\cos\theta.
\]
The second factor is an expression of
the well-known result that the degeneracy is
broken nonperturbatively; the energies form a continuum, depending on
the value of $\theta$.

Another application of instantons in quantum mechanics is the
phenomenon of tunneling (barrier penetration). The instanton method
can be used to calculate the lifetime of a metastable state in a
potential of the form depicted in Figure 15. We will not discuss this
application. 

Instantons also appear in (and are by far most useful in) field
theory. In certain field theories the space of finite-Euclidean-action
configurations separates into distinct topological classes. An
instanton is a nontrivial configuration of this type. The necessary
topological requirements for this to occur are not hard to satisfy,
and the list of theories that have instantons includes the Abelian
Higgs model in 1+1 dimensions, the O(3) nonlinear $\si$-model in 1+1
dimensions, the Skyrme model in 2+1 dimensions, and (most
significantly)
QCD. Instantons give rise to a host of interesting phenomena depending
on the model, including confinement (not in QCD though!),
$\theta$-vacua, a solution of the U(1) problem in strong
interactions, and the decay of a metastable vacuum. Unfortunately time
does not permit discussion of these fascinating phenomena.

\newpage\thispagestyle{empty}
\section[Summary]{Summary and Gross Omissions}

In this set of lectures the subject of path integrals has been covered
starting from scratch, emphasizing explicit calculations in quantum
mechanics. This emphasis has its price: I have not had time to cover
several things I would have liked to discuss.
My hope is that having been subjected to calculations
in gory detail for the most part
in the relatively familiar context of quantum mechanics, you will be
able to study  more complicated and interesting
applications on your own.

Here is a list of some of the important aspects and applications of
this subject which I didn't have time to discuss:
\ben
\item Fermi fields and Grassmann functional integration;
\item Gauge theories (gauge fixing and ghosts arise in a particularly
  elegant way);
\item Feynman's variational method and application to the polaron
  (electron moving in a crystal environment);
\item Derivation of the Landau-Ginsburg theory, including application
  to superconductivity;
\item Instantons in field theory;
\item Critical phenomena.
\een

I hope that, in spite of these unforgivable omissions, these lectures
have been worthwhile.

\section{Acknowledgements}

I wish to thank the Rencontres du Vietnam and the NCST, Hanoi for the
invitation to give these lectures, and especially Patrick Aurenche for
his boundless energy before, during, and after the School. I am
grateful to Jean-S\'ebastien Caux and Manu Paranjape for discussions
and advice during the preparation of these lectures, and to Patrick
Irwin, who spotted many weaknesses and errors in this manuscript after
I had considered it fit for human consumption.

This work was financed in part by the Natural Sciences and Engineering
Research Council of Canada.

\newpage\thispagestyle{empty}
\section{References}

\ben
\item P.A.M. Dirac, Physikalische Zeitschrift der Sowjetunion, Band 3,
  Heft 1 (1933).
\item R.P. Feynman, Reviews of Modern Physics 20, 367 1948.
\item R.P. Feynman and A.R. Hibbs, {\sl Quantum Mechanics and Path
    Integrals}, McGraw-Hill, 1965.
\item R.P. Feynman, {\sl Statistical Mechanics: A Set of Lectures},
  Benjamin, 1972.
\item G. Baym, {\sl Lectures on Quantum Mechanics}, Benjamin-Cummings,
  1973.
\item L.S. Schulman, {\sl Techniques and Applications of Path
    Integration}, John Wiley and Sons, 1981.
\item D.J. Amit, {\sl Field Theory, the Renormalization Group and
  Critical Phenomena}, 2nd Edition, World Scientific, 1984.
\item L.H. Ryder, {\sl Quantum Field Theory}, Cambridge University
  Press, 1985.
\item P. Ramond, {\sl Field Theory: A Modern Primer, Second Edition},
  Addison-Wesley, 1990. 
\item E.S. Abers and B.W. Lee, Physics Reports 9C, 1, 1973.
\item B. Sakita, {\sl Quantum Theory of Many-Variable Systems and
  Fields}, World Scientific, 1985.
\item S. Coleman, {\sl Aspects of Symmetry}, Cambridge University
  Press, 1985.
\item A.M. Tsvelik, {\sl Quantum Field Theory in Condensed Matter
    Physics}, Cambridge University Press, 1995.
\item V.N. Popov, {\sl Functional Integrals and Collective
    Excitations}, Cambridge University Press, 1987.
\item E. Fradkin, {\sl Field Theories of Condensed Matter Systems},
  Addison-Wesley, 1991.
\een

\end{document}